\documentclass[reqno,11pt]{amsart}
\usepackage{amssymb, amsmath}
\usepackage{amsthm, amsfonts,mathrsfs}
\usepackage{BOONDOX-cal}
\usepackage[T1]{fontenc}
\usepackage{times}
\usepackage{graphicx}
\usepackage{epsfig}
\usepackage{color}
\usepackage[all]{xypic}
\usepackage{setspace}
\newtheorem{remark}{Remark}[section]

\numberwithin{equation}{section}

\newcommand{\mr}{\mathbb{R}}

\newcommand{\normmm}[1]{{\left\vert\kern-0.25ex\left\vert\kern-0.25ex\left\vert #1\right\vert\kern-0.25ex\right\vert\kern-0.25ex\right\vert}}
\allowdisplaybreaks[4]

\begin{document}
\begin{spacing}{1.5}
\title[Coupled surface and internal waves over currents and uneven bottom]
{Hamiltonian model for coupled
surface and internal waves over currents and uneven bottom}%
\author[Lili Fan]{Lili Fan}%
\address[Lili Fan]{College of Mathematics and Information Science,
Henan Normal University, Xinxiang 453007, China}
\email{fanlily89@126.com}
\author[Ruonan Liu]{Ruonan Liu}%
\address[Ruonan Liu]{College of Mathematics and Information Science,
Henan Normal University, Xinxiang 453007, China}
\email{liuruonan97v@163.com}
\author[Hongjun Gao $^{\dag}$]{Hongjun Gao $^{\dag}$}
\address[Hongjun Gao]{School of Mathematics, Southeast University, Nanjing 211189, PR China}
\email{hjgao@seu.edu.cn\, (Corresponding author)}


\begin{abstract}
A Hamiltonian model for the propagation of internal water waves interacting with surface waves, a current and an uneven bottom is examined. Using the so-called Dirichlet-Neumann operators, the water wave system is expressed in the Hamiltonian form, and thus the motions of the internal waves and surface waves are determined by the Hamiltonian formulation. Choosing an appropriate scaling of the variables and employing the Hamiltonian perturbation theory from Hamiltonian formulation of the dynamics,  we derive a KdV-type equation with variable coefficients depending on the bottom topography to describe the internal waves.
\end{abstract}

\date{}

\maketitle




\section{Introduction}
Consideration in this paper is a water wave system describing the two-dimensional nonlinear interaction between coupled surface waves, internal waves, and an underlying current with piecewise constant vorticity, in a two-layered fluid overlying an uneven bottom. Internal waves can occur in the interior stratified region of strong temperature or density gradients (thermocline or pycnocline) in the sea. On account of the physical significance of the internal waves in the analysis of the energy revenue and expenditure and the energy balance in the ocean dynamics and their significant applications in the exploitation and protection of the ocean environment and resources, there are by now numerous studies on internal waves (e.g. \cite{BB1,BB2,CC,CC1}). The consideration of current to our water wave system is due to its significant mathematical and physical features and complexity in geophysical dynamics. Among many interested groups, the interaction of nonlinear waves and currents is of particular interest to oceanographers and climatologists due to its applications for the prediction of tsunami, rogue waves, etc. \cite{Co,CoJ15,CoJ2017,CoM}. The problem of waves with a variable bottom can be referred to the pioneering work of Johnson \cite{Jo,Jo1}, where a perturbed Korteweg-de Vries (KdV) equation \cite{KdV} is derived as a model for surface waves from Euler's governing equations for irrotational inviscid fluid. Given all the above consideration, a system of coupled surface waves, internal waves, with depth-dependent currents in each domain and over an uneven bottom is of particular interest both mathematically and physically.

It is known that linear approximations are the usual means to make predictions of water-wave propagation in oceanography. Although this approach is successful in many instances, for complex flow patterns an adequate description of the phenomenon can not neglect nonlinear effects. Motivated by the recent works \cite{CoI19,IMT}, we develop a nonlinear approach that captures the main features of the dynamics under consideration by adopting the Hamiltonian framework to manage our system, which is also amenable to approximations in shallow-water (long wave) regime. The Hamiltonian perspective to deal with water waves can date back to Zakharov's pioneering work \cite{Za}, where he found that the governing equations for fluids of inviscidity, irrotationality, incompressibility and uniformity of density form a canonical Hamiltonian formulation. This opened a novel way to explore the Hamiltonian canonical structure of the problems on water waves.  Subsequently, this approach has been extended successfully to deal with two-layer flows \cite{CG,CGK,D,LDZ}. The Hamiltonian perspective to handle models with finite depth and shear with constant vorticity can be referred to \cite{CoIP,Wa}, and for wave-current interactions in a two-layer system we refer to \cite{CoI15,CoI19,CoIM,C1,C2,CI,CI0,CI1,CuI,IM,I,M} and the references therein. Meanwhile, the Hamiltonian approach to the wave motion by the inclusion of variations of the bottom surface can be referred to the recent works of \cite{CIMT,CIT,CGNS,DCDPG} for single layer flows and of \cite{IMT,ZP} for two-layer fluids.  The purpose of the present paper is to extend the results of Refs. \cite{CoI19,IMT} to the cases of coupled surface and internal waves with variable bottom.

To make it convenient to analyse, we first determine the formulation of the problem in terms of Hamiltonian with the aid of the Dirichlet-Neumann operators and express the system in terms of canonical wave-related variables. The appearance of the bottom function complicates the obtained Hamiltonian structure, which is a starting point for the derivation of simpler models. Using the Hamiltonian perturbation theory \cite{CGK,CGNS,CGS} with a specific scaling of the variables inspired by \cite{CoI19}, we develop a systematic perturbation analysis on the obtained Hamiltonian formulation to obtain the surface and internal wave motion equations in the shallow-water (long wave) regime, where the intractable Taylor expansions of the Dirichlet-Neumann operators, simpler ones compared to \cite{CI1}, are the core. These simplified equations enable an in-depth study of the coupling between the surface and internal waves, and how both these wave systems interact with the background current and variable bottom.

Furthermore, we derive the KdV-type equation for the internal waves. The obtained effective KdV-type equations with variable coefficients depending on the bottom topography recover the corresponding ones in \cite{CoI19} for fluid without the consideration of bottom variations and the dispersion relation recovers the results in \cite{CIMT} for single-layer system. Our methods for a system with both surface and internal waves over uneven bottom answers partially the problems proposed in \cite{CIMT} and we hope the study being undertaken enable an in-depth study in these circumstances.

The remainder of this paper is organized as follows.  In Section 2, we present the flows we study and give the governing equations for the problem. In Section 3, we propose the Hamiltonian formulation of the system understudied and a systematic long-wave perturbation analysis by the Hamiltonian perturbation theory is given in Section 4. Section 5 focuses on the derivation of the KdV-type equation describing the evolution of the internal waves.
\section{Equations of motion}

\subsection{System set-up}
\large
The fluid system consists of the point $(x, y)$ such that $B(x)< y < h_2$, and it is divided into two regions
\begin{align*}
&\Omega_1(t;\eta,B) = \{(x, y)\in\mr^2 : B(x) < y < \eta(x,t)\} \\
\text{and} \qquad &\Omega_2(t; \eta,\eta_2) = \{(x, y) : \eta(x,t) < y < h_2+\eta_2(x,t)\}.
\end{align*}
Here $B(x):=-h_1+\beta(x)$ is the stationary impermeable bottom for $-h_1$ is the average bottom level and $\beta(x)$ is the bottom elevation function, $\eta(x,t)$ denotes the interface elevation and the undisturbed interface is chosen at $y = 0$, $h_2+\eta_2(x,t)$ is a surface wave around the average level $y=h_2$. The subscript notations $1$, $2$ refer to values in $\Omega_1$ and $\Omega_2$ respectively and $i = \{1; 2\}$ for values in both domains.
\begin{figure}[htbp]
\centering\includegraphics[width=4in]{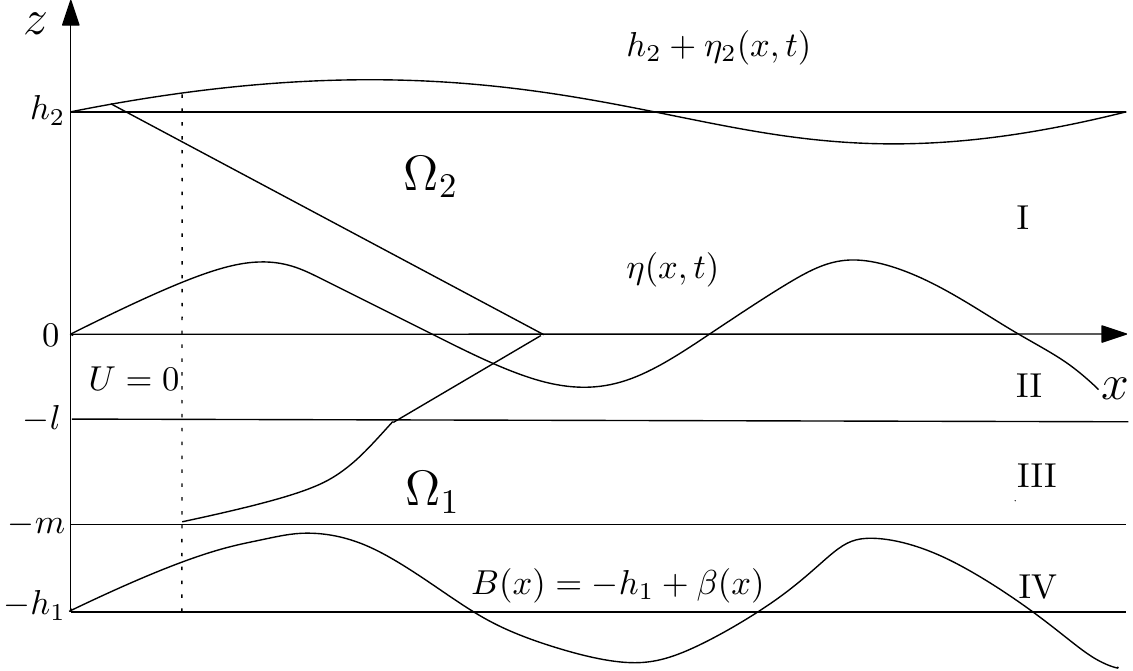}
\caption{System with an internal wave over uneven bottom}\label{fig1}
\end{figure}
The average values for the interface elevation, surface elevation and the bottom level are assumed to be zero,
\begin{equation}\label{2.1}
\int_{\mathbb{R}} \eta(x, t) d x=0, \quad \int_{\mathbb{R}} \eta_2(x, t) d x=0, \quad \int_{\mathbb{R}} \beta(x, t) d x=0.
\end{equation}
Suppose that the two regions are occupied by two homogeneous, inviscid and incompressible fluids with density $\rho_2$ of the upper fluid and $\rho_1$ of the lower fluid. The stable configuration is given by the immiscibility condition $\rho_1>\rho_2$.

We define the stream functions, $\psi_i$, as
\begin{equation}\label{2.2}
u_{i}=\psi_{i, y} \;\;\text {and} \;\; v_{i}=-\psi_{i, x},
\end{equation}
and velocity potentials, $\varphi_i$, as
\begin{equation}\label{2.3}
u_{i}=\varphi_{i, x}+\gamma_{i} y \;\; \text { and } \;\; v_{i}=\varphi_{i, y},
\end{equation}
where $\mathbf{u}_{i}=\left(u_{i}, v_{i}, 0\right)$ denote the velocity fields and $\gamma_{i}=-v_{i,x}+u_{i,y}$ represent the constant vorticities. This setting allows for modelling of an undercurrent, such as the Equatorial Undercurrent, and \eqref{2.3} can be written as \cite{CoI15}
\begin{equation}\label{2.4}
u_{i}=\widetilde{\varphi}_{i, x}+U(y)\;\; \text { and }\;\; v_{i}=\widetilde{\varphi}_{i, y},
\end{equation}
where $U(y)$ is the current profile defined by
\begin{equation}\label{2.5}
U(y)=\begin{cases}
\gamma_{2} y+\kappa   \quad\;\text { for } \eta(x,t) \leq y \leq h_2+\eta
_2(x,t)  \quad \text { (layer I) } \\
\gamma_{1} y+\kappa  \quad\;\text { for } -l\leq y \leq \eta(x,t)\quad \text { (layer II) } \\
U_{1}(y)   \qquad\text{ for }-m \leq y \leq -l \quad \text { (layer III) } \\
0   \qquad\qquad\text { for } B(x) \leq y \leq -m \quad \text { (layer IV) }.
\end{cases}
\end{equation}
Here the tilde notation is introduced to separate out the wave-only components, $\kappa$ is a constant component of the current, $ U_1(y) $ is a continuous function such that $ U_1(-m)=0, U_1(-l)=-l \gamma_{1}+\kappa $ and the current has a jump when $ \eta \neq 0 $ as illustrated on Fig. 1 for the situation of an undisturbed fluid when $ \eta \equiv 0 $.

We make the assumption that, for any $y$ and $t$, the functions $\eta(x, t)$, $\eta_2(x, t)$,  $\widetilde{\varphi}_{i}(x, y, t)$ and $\beta(x)$ are in the Schwartz class $\mathcal{S}(\mathbb{R})$ with respect to $x$, where the Schwartz class $\mathcal{S}(\mathbb{R})$ denotes the smooth, rapidly decreasing functional space. We make the assumptions
\begin{align}
&\lim _{|x| \rightarrow \infty} \eta(x, t)=0, \qquad
\lim _{|x| \rightarrow \infty} \eta_2(x, t)=0, \label{2.6}\\
&\lim _{|x| \rightarrow \infty} \beta(x)=0,\qquad
\lim _{|x| \rightarrow \infty} \tilde{\varphi}_{i}(x, y, t)=0, \label{2.7}
\end{align}
and the condition that the stream function is continuous on the interface is given by
\begin{equation}\label{2.8-1}
\psi_{1}(x, \eta, t)=\psi_{2}(x, \eta, t).
\end{equation}
\subsection{Governing equations}
The Euler equations for the two layers are
\begin{equation}\label{2.8}
\mathbf{u}_{i, t}+\left(\mathbf{u}_{i} \cdot \nabla\right) \mathbf{u}_{i}=-\frac{1}{\rho_{i}} \nabla p_{i}+\mathbf{g}+\mathbf{F}_{i},
\end{equation}
where $\mathbf{F}_{i}=2 \omega \nabla \psi_{i}$ are the Coriolis forces per unit mass at the equator with $\omega$ being the rotational speed of the Earth, $\mathbf{g}=(0,0,-g)$ with $g$ being the gravitational acceleration, and $p_{i}$ are the corresponding pressures. The equation of mass conservation for incompressible fluid is given by
\begin{equation}\label{2.9}
u_{i,x}+v_{i,y}=0.
\end{equation}
The kinematic boundary condition at the interface is given as
\begin{equation}\label{2.12}
\eta_{t}=v_{i}-u_{i} \eta_{x},
\end{equation}
which, when using \eqref{2.4} of the velocity potentials, can be expressed as
\begin{equation}\label{2.13}
\eta_{t}=\left(\widetilde{\varphi}_{i, y}\right)_{c}-\left(\left(\widetilde{\varphi}_{i, x}\right)_{c}+\gamma_{i} \eta+\kappa\right) \eta_{x}.
\end{equation}
The kinematic boundary condition on the bed $y=B(x)$ is given by
\begin{equation}\label{2.14}
v_1(x, B(x), t)=u_1(x, B(x), t) B^{\prime}(x), \;\; \text{or} \;\; \left(\widetilde{\varphi}_{1, y}\right)_{b}=\left(\widetilde{\varphi}_{1, x}\right)_{b} B^{\prime}(x),
\end{equation}
where the subscript notation $b$ refers to values at the bottom, and, additionally, the kinematic boundary condition at the top $y=h_2+\eta_2$ is given by
\begin{equation}\label{2.15}
\eta_{2,t}=\left(\widetilde{\varphi}_{2, y}\right)_{t}-\left(\left(\widetilde{\varphi}_{2, x}\right)_{t}+\gamma_{2} (h_2+\eta_2)+\kappa\right) \eta_{2,x}.
\end{equation}

By \eqref{2.2}, \eqref{2.4}, \eqref{2.8} and \eqref{2.9}, we can express the pressure gradients as
\begin{equation}\label{2.10}
\nabla p_{i}=-\rho_{i} \nabla\left(\tilde{\varphi}_{i, t}+\frac{1}{2}\left|\nabla \psi_{i}\right|^{2}-\left(\gamma_{i}+2 \omega\right) \psi_{i}+g y\right).
\end{equation}
Moreover, considering the dynamic boundary condition at the interface $p_{1}=p_{2}$, one can obtain the Bernoulli condition \cite{I}
\begin{align}\label{2.11}
&\rho_{1}\left(\left(\widetilde{\varphi}_{1, t}\right)_{c}+\frac{1}{2}\left|\nabla \psi_{1}\right|_{c}^{2}-(\gamma_{1}+2 \omega) \chi+g \eta\right)\nonumber\\
&\quad=\rho_{2}\left(\left(\tilde{\varphi}_{2, t}\right)_{c}+\frac{1}{2}\left|\nabla \psi_{2}\right|_{c}^{2}-\left(\gamma_{2}+2 \omega\right) \chi+g \eta\right),
\end{align}
where the subscript $c$ is introduced to signify evaluation at the common interface $y=\eta(x, t)$ and
\begin{equation}\label{6.32}
\chi=\psi_{1}(x, \eta, t)=\psi_{2}(x, \eta, t) = - \int _ { - \infty } ^ { x } \eta _ { t } \left( x ^ { \prime } , t \right) d x ^ { \prime },
\end{equation}
by \eqref{2.8-1} and
\begin{equation*}
\frac { d } { d x } \chi =\frac { d } { d x } \psi _ { i } ( x , \eta ( x , t ) , t )
=\left( \psi _ { i , x } \right) _ { c } + \left( \psi _ { i , y } \right) _ { c } \eta _ { x }=-\eta_t.
\end{equation*}
The dynamic boundary condition at the surface $p_{2}=Const.$ yields that \cite{I}
\begin{equation}\label{2.11-1}
(\tilde{\varphi}_{2, t})_{s}+\frac{1}{2}(\left|\nabla \psi_{2}\right|^{2})_{s}-(\gamma_{2}+2 \omega) \chi_{2}+g\left(h_{2}+\eta_{2}\right)=0,
\end{equation}
where the subscript $s$ refers to values at the surface $y=h_2+\eta_2$ and $\chi_{2}=\psi_{2}(x, h_2+\eta_2, t)$.
\section{Hamiltonian formulation}
\subsection{Hamiltonian formulation by means of the Dirichlet-Neumann operator}
Following the idea in \cite{CoI19,CIT,CGNS,CGS}, we present the Hamiltonian formulation for \eqref{2.8}-\eqref{2.15} in this section. The Hamiltonian of the general non-flat bottom system is given by
\begin{align}\label{6.1}
&H\left(\eta, \mathbf{u}_{i}\right)=\frac{1}{2} \rho_{1} \int_{\mathbb{R}} \int_{B}^{\eta}\left(u_{1}^{2}+v_{1}^{2}\right) d y d x+\frac{1}{2} \rho_{2} \int_{\mathbb{R}} \int_{\eta}^{h_{2}+\eta_{2}}\left(u_{2}^{2}+v_{2}^{2}\right) d y d x \nonumber\\
&\qquad\qquad\;\;\;+\rho_{1} g \int_{\mathbb{R}} \int_{B}^{\eta} y dy d x+\rho_{2} g \int_{\mathbb{R}} \int_{\eta}^{h_{2}+\eta_{2}} y dy d x+\int_{\mathbb{R}} \mathfrak{h}_{0} d x,
\end{align}
which is the sum of the kinetic and potential energies. The notation $\mathfrak{h}_{0}$ is a constant Hamiltonian density. By \eqref{2.4} of the relationship between $\mathbf{u}_{i}$ and $\widetilde{\varphi}_{i}$, we obtain
\begin{equation}\label{6.2}
u_{i}^{2}+v_{i}^{2}=\left|\nabla \widetilde{\varphi}_{i}\right|^{2}+2 U \widetilde{\varphi}_{i, x}+U^{2},
\end{equation}
and hence the Hamiltonian \eqref{6.1} can be written in terms of $\eta(x, t)$ and
$\tilde{\varphi}_{i}(x, t)$ as
\begin{align}\label{6.3}
&H\left(\eta, \widetilde{\varphi}_{i}\right)
=\frac{1}{2} \rho_{1} \int_{\mathbb{R}} \int_{B}^{\eta}\left|\nabla \widetilde{\varphi}_{1}\right|^{2} d y d x+\frac{1}{2} \rho_{2} \int_{\mathbb{R}} \int_{\eta}^{h_{2}+\eta_{2}}\left|\nabla \widetilde{\varphi}_{2}\right|^{2} d y d x \nonumber\\
&+\rho_{1} \int_{\mathbb{R}} \int_{B}^{\eta} U \widetilde{\varphi}_{1, x} d y d x+\rho_{2} \int_{\mathbb{R}} \int_{\eta}^{h_{2}+\eta_{2}} U \widetilde{\varphi}_{2, x} d y d x+\frac{1}{2} \rho_{1} \int_{\mathbb{R}} \int_{B}^{\eta} U^{2} d y d x \nonumber\\
&+\frac{1}{2} \rho_{2} \int_{\mathbb{R}} \int_{\eta}^{h_{2}+\eta_2} U^{2} d y d x+\rho_{1} g \int_{\mathbb{R}} \int_{B}^{\eta} y d y d x+\rho_{2} g \int_{\mathbb{R}} \int_{\eta}^{h_{2}+\eta_{2}} y d y d x+\int_{\mathbb{R}} \mathfrak{h}_{0} d x.
\end{align}
By the definition of $U(y)$ in \eqref{2.5}, we split the current into 4 layers to reach that
\begin{align}\label{6.4}
&H\left(\eta, \widetilde{\varphi}_{i}\right)\nonumber\\
&=\frac{1}{2} \rho_{1} \int_{\mathbb{R}} \int_{B}^{\eta}\left|\nabla \widetilde{\varphi}_{1}\right|^{2} d y d x+\frac{1}{2} \rho_{2} \int_{\mathbb{R}} \int_{\eta}^{h_{2}+\eta_{2}}\left|\nabla \widetilde{\varphi}_{2}\right|^{2} d y d x+\rho_{1} \int_{\mathbb{R}} \int_{-m}^{-l} U_{1} \tilde{\varphi}_{1, x} d y d x \nonumber\\
&+\rho_{1} \gamma_{1} \int_{\mathbb{R}} \int_{-l}^{\eta} y \widetilde{\varphi}_{1, x} d y d x+\rho_{2} \gamma_{2} \int_{\mathbb{R}} \int_{\eta}^{h_2+\eta_{2}} y \widetilde{\varphi}_{2, x} d y d x+\rho_{1} \kappa \int_{\mathbb{R}} \int_{-l}^{\eta} \tilde{\varphi}_{1, x} d y d x\nonumber\\
&+\rho_{2} \kappa \int_{\mathbb{R}} \int_{\eta}^{h_2+\eta_{2}} \tilde{\varphi}_{2, x} d y d x+\frac{1}{2} \rho_{1} \int_{\mathbb{R}} \int_{-m}^{-l} U_{1}^{2} d y d x
+\frac{1}{2} \rho_{2} \int_{\mathbb{R}} \int_{\eta}^{h_2+\eta_{2}}\left(\gamma_{2} y+\kappa_{2}\right)^{2} d y d x\nonumber\\
&+\frac{1}{2} \rho_{1} \int_{\mathbb{R}} \int_{-l}^{\eta}\left(\gamma_{1} y+\kappa\right)^{2} d y d x+\rho_{1} g \int_{\mathbb{R}} \int_{B}^{\eta} y d y d x+\rho_{2} g \int_{\mathbb{R}} \int_{\eta}^{h_2+\eta_{2}} y d y d x+\int_{\mathbb{R}} \mathfrak{h}_{0} d x.
\end{align}
In order to obtain the Hamiltonian in terms of variables defined at the interface and surface, we introduce the notations for the velocity potentials values at the interface and surface as
\begin{equation}\label{6.5}
\phi_{i}:=\left(\tilde{\varphi}_{i}\right)_{c},\quad \phi_{3}:=\left(\tilde{\varphi}_{2}\right)_{s}=\tilde{\varphi}_{2}(x,h_2+\eta_2,t),
\end{equation}
and the so-called Dirichlet-Neumann (DN) operators \cite{CGNS,CGS}
\begin{align}
&G_{1}(\beta,\eta) \phi_{1}=-\eta_{x}\left(\tilde{\varphi}_{1, x}\right)_{c}+\left(\tilde{\varphi}_{1, y}\right)_{c}=\left(\nabla \tilde{\varphi}_{1}\right)_{c }\cdot\left(\mathbf{n}_{1}\right)_{c} \sqrt{1+\eta_{x}^{2}}, \label{6.6}\\
& G_{2}(\eta, \eta_2) \phi_{2}=\eta_{x}\left(\widetilde{\varphi}_{2, x}\right)_{c}-\left(\widetilde{\varphi}_{2, y}\right)_{c}=\left(\nabla \tilde{\varphi}_{2}\right)_{c }\cdot\left(\mathbf{n}_{2}\right)_{c} \sqrt{1+\eta_{x}^{2}}, \label{6.7}\\
\text { and } & \quad G_{2}(\eta, \eta_2) \phi_{3}=-\eta_{2,x}\left(\tilde{\varphi}_{2, x}\right)_{t}+\left(\tilde{\varphi}_{2, y}\right)_{t}=\left(\nabla \tilde{\varphi}_{2}\right)_{s}\cdot\left(\mathbf{n}_{3}\right)_{s} \sqrt{1+\eta_{2,x}^{2}}, \label{6.6-1}
\end{align}
where $\mathbf{n}_{i}$ for $i=1,2,3$ are the unit exterior normal. The entries of the matrix operator $G_{2}(\eta, \eta_2)$ are given by
\begin{equation}\label{G}
G_{2}\left(\eta, \eta_{2}\right)
=\begin{pmatrix}G_{11} & G_{12} \\ G_{21} & G_{22}
\end{pmatrix}.
\end{equation}
Using
\begin{equation}\label{6.8}
\left|\nabla \tilde{\varphi}_{i}\right|^{2}=\nabla \cdot\left(\left(\nabla \tilde{\varphi}_{i}\right) \tilde{\varphi}_{i}\right)=\operatorname{div}(\tilde{\varphi}_{i} \nabla \tilde{\varphi}_{i})
\end{equation}
and the divergence theorem, the following $\Omega_{1}$ integral can be expressed as
\begin{equation*}
\begin{aligned}
&\int_{\mathbb{R}}\int_{B}^{\eta} \nabla \cdot\left(\left(\nabla \widetilde{\varphi}_{1}\right) \tilde{\varphi}_{1}\right) d y d x \\
&\quad=\int_{\mathbb{R}}\left(\left(\nabla \widetilde{\varphi}_{1}\right)_{c} \phi_{1}\right) \cdot\left(\mathbf{n}_{1}\right)_{c} \sqrt{1+\left(\eta_{x}\right)^{2}} d x+\int_{\mathbb{R}}\left(\left(\nabla \widetilde{\varphi}_{1}\right)_{b}\left(\widetilde{\varphi}_{1}\right)_{b}\right) \cdot\left(\mathbf{n}_{1}\right)_{b} \sqrt{1+\left(\eta_{x}\right)^{2}} d x.
\end{aligned}
\end{equation*}
Noting that
\begin{equation*}
\left(\mathbf{n}_{1}\right)_{c}=\left(-\eta_{x}, 1\right) \text { and }\left(\mathbf{n}_{1}\right)_{b}=(B'(x),-1)
\end{equation*}
and from \eqref{2.14}
\begin{equation*}
\left(\nabla \tilde{\varphi}_{1}\right)_{b} \cdot\left(\mathbf{n}_{1}\right)_{b}=\left(\left(\tilde{\varphi}_{1, x}\right)_b, \left(\tilde{\varphi}_{1, y}\right)_b\right) \cdot(B'(x),-1)=0,
\end{equation*}
give that
\begin{align}\label{6.9}
\frac{1}{2} \rho_{1} \int_{\mathbb{R}}\int_{B}^{\eta} \nabla \cdot\left(\left(\nabla \widetilde{\varphi}_{1}\right) \tilde{\varphi}_{1}\right) d y d x
&=\frac{1}{2} \rho_{1}\int_{\mathbb{R}}\left(\left(\nabla \widetilde{\varphi}_{1}\right)_{c} \phi_{1}\right) \cdot\left(\mathbf{n}_{1}\right)_{c} \sqrt{1+\left(\eta_{x}\right)^{2}} d x\nonumber\\
&=\frac{1}{2} \rho_{1}\int_{\mathbb{R}} \phi_{1} G_{1}(\beta,\eta) \phi_{1} d x.
\end{align}
Similarly,
\begin{equation}\label{6.10}
\frac{1}{2} \rho_{2} \int_{\mathbb{R}} \int_{\eta}^{h_{2}+\eta_2}\left|\nabla \tilde{\varphi}_{2}\right|^{2} d y d x=\frac{1}{2} \rho_{2} \int_{\mathbb{R}}\left( \phi_{3} G_{2}(\eta,\eta_2) \phi_{3}+\phi_{2} G_{2}(\eta,\eta_2) \phi_{2} \right)d x.
\end{equation}
Employing the kinematic boundary conditions from \eqref{2.13}
\begin{align}
&G_{1}(\beta,\eta) \phi_{1}=-\eta_{x}\left(\widetilde{\varphi}_{1, x}\right)_{c}+\left(\widetilde{\varphi}_{1, y}\right)_{c}=\eta_{t}+\left(\gamma_{1} \eta+\kappa\right) \eta_{x}, \label{6.11}\\
&G_{11} \phi_{2}+G_{12} \phi_{3}=\eta_{x}\left(\widetilde{\varphi}_{2, x}\right)_{c}-\left(\widetilde{\varphi}_{2, y}\right)_{c}=-\eta_{t}-\left(\gamma_{2} \eta+\kappa\right) \eta_{x},\label{6.12}\\
&
G_{21} \phi_{2}+G_{22} \phi_{3}=-\eta_{2,x}\left(\widetilde{\varphi}_{2, x}\right)_{t}+\left(\widetilde{\varphi}_{2, y}\right)_{t}=\eta_{2,t}+\left(\gamma_{2}h_2+\gamma_{2} \eta_2+\kappa\right) \eta_{2,x},\label{6.13}
\end{align}
we can obtain that
\begin{equation}\label{6.14}
G_{1}(\beta,\eta) \phi_{1}+G_{11} \phi_{2}+G_{12} \phi_{3}=(\gamma_{1}-\gamma_{2})\eta\eta_x.
\end{equation}
The variables
\begin{equation}\label{6.15}
\xi:=\rho_{1} \phi_{1}-\rho_{2} \phi_{2},\quad \xi_2:=\rho_{2}\phi_{3}
\end{equation}
are introduced as in \cite{BB1,BB2} to serve as generalised momentum. Then we can write
\begin{equation*}
\begin{aligned}
&\left(\rho_{1} G_{11}+\rho_{2} G_{1}(\beta,\eta)\right) \phi_{1}=G_{11} \xi-G_{12} \xi_2+\rho_{2} (\gamma_{1}-\gamma_{2})\eta\eta_x \\
\text { and } \quad&\left(\rho_{1} G_{11}+\rho_{2} G_{1}(\beta,\eta)\right) \phi_{2}=-G_{1}\xi-\frac {\rho_1} {\rho_2}G_{12}\xi_2+\rho_{1} (\gamma_{1}-\gamma_{2})\eta\eta_x .
\end{aligned}
\end{equation*}
We need also the operator
\begin{equation}\label{6.16}
\mathcal{B}(B,\eta,\eta_2):=\rho_{1} G_{11}+\rho_{2} G_{1},
\end{equation}
and the potentials $\phi_{1}$, $\phi_{2}$ and $\phi_{3}$ can be solved as
\begin{align}
\phi_{1} &=\mathcal{B}^{-1}\left(G_{11} \xi-G_{12} \xi_2+\rho_{2} (\gamma_{1}-\gamma_{2})\eta\eta_x\right), \label{6.17}\\
\phi_{2} &=\mathcal{B}^{-1}\left(-G_{1}\xi-\frac {\rho_1} {\rho_2}G_{12}\xi_2+\rho_{1} (\gamma_{1}-\gamma_{2})\eta\eta_x \right), \label{6.18}\\
\phi_{3} &=\frac 1 {\rho_{2}} \xi_2. \label{6.18-1}
\end{align}
So the first two terms of \eqref{6.4} can be written as
\begin{align}\label{6.19}
&\frac{1}{2} \rho_{1} \int_{\mathbb{R}} \int_{B}^{\eta}\left|\nabla \tilde{\varphi}_{1}\right|^{2} d y d x+\frac{1}{2} \rho_{2} \int_{\mathbb{R}} \int_{\eta}^{h_{2}+\eta_2}\left|\nabla \tilde{\varphi}_{2}\right|^{2} d y d x \nonumber\\
&=\frac{\rho_{1}}{2}\int_{\mathbb{R}} \phi_{1} \left(\eta_{t}+\gamma_1 \eta \eta_{x}+\kappa \eta_{x}\right) d x\nonumber\\
&\quad+\frac{\rho_{2}}{2}\int_{\mathbb{R}}
\begin{pmatrix}\phi_{2} & \phi_{3}\end{pmatrix}
\begin{pmatrix}{c}-\eta_{t}-\gamma_{2} \eta \eta_{x}-\kappa\eta_{x} \\
\eta_{2, t}+\gamma_{2}\left(\eta_{2}+h_{2}\right) \eta_{2, x}+\kappa\eta_{2, x}\end{pmatrix}
 d x\nonumber\\
&=\frac{1}{2} \int_{\mathbb{R}} \left[\xi \eta_{t}+\xi_{2}\left(\eta_{2, t}+\left(\gamma_{2}\left(\eta_{2}+h_{2}\right)+\kappa\right) \eta_{2, x}\right)
+\left(\rho_1 \gamma_1 \phi_1-\rho_{2} \gamma_{2} \phi_{2}\right) \eta \eta_{x}+\xi \kappa \eta_{x}\right]d x\nonumber\\
&=\frac{1}{2} \int_{\mathbb{R}} \left[\begin{pmatrix}\xi \\ \xi_{2}\end{pmatrix}^{T}\begin{pmatrix}\eta_{t}+\kappa\eta_x \\ \eta_{2, t}+\left(\gamma_{2}\left(\eta_{2}+h_{2}\right)+\kappa\right) \eta_{2, x}\end{pmatrix}  +\left(\rho_1 \gamma_1 \phi_1-\rho_{2} \gamma_{2} \phi_{2}\right) \eta \eta_{x} \right]d x \nonumber\\
&=\frac{1}{2} \int_{\mathbb{R}}  \begin{pmatrix}\xi \\ \xi_{2}\end{pmatrix}^{T}\begin{pmatrix}-G_{11} \phi_{2}-G_{12} \phi_{3}-\gamma_{2} \eta \eta_{x} \\ G_{21} \phi_{2}+G_{22} \phi_{3}\end{pmatrix}  d x
+\frac{1}{2}\int_{\mathbb{R}} \left(\rho_1 \gamma_1 \phi_1-\rho_{2} \gamma_{2} \phi_{2}\right) \eta \eta_{x} d x\nonumber\\
&=\frac{1}{2} \int_{\mathbb{R}} \begin{pmatrix}\xi \\ \xi_{2}\end{pmatrix}^{T}\begin{pmatrix}-G_{11} & -G_{12} \\ G_{21} & G_{22}\end{pmatrix}\begin{pmatrix}\phi_{2} \\ \phi_{3}\end{pmatrix} d x\nonumber\\
&\quad+\frac{1}{2}\int_{\mathbb{R}} \left(\rho_1 \gamma_1 \phi_1-\rho_{2} \gamma_{2} \phi_{2}\right) \eta \eta_{x} d x-\frac{1}{2}\int_{\mathbb{R}} \gamma_{2} \eta \eta_{x}\xi d x.
\end{align}

The third term of \eqref{6.4} is zero due to the fact that $\int _ { \mathbb { R } } \tilde { \varphi } _ { 1 , x } d x = 0$. Now we turn to the fourth and fifth terms of \eqref{6.4}. As
\begin{align*}
&\int _ { \mathbb { R } }\int _ { - l} ^ { \eta ( x ) } y \widetilde { \varphi } _ { 1 , x } d ydx = - \int _ { \mathbb { R } }\phi _ { 1 } \eta \eta _ { x }dx,\\
&\int _ { \mathbb { R } }\int _ { \eta ( x )} ^ { h _ { 2 } +\eta_2 } y \widetilde { \varphi } _ { 2 , x } d ydx =  \int _ { \mathbb { R } }\phi _ { 2 } \eta \eta _ { x }dx-\int _ { \mathbb { R } }\phi _ { 3 } (\eta_2+h_2) \eta _ { 2, x }dx,
\end{align*}
and noting the expressions for $\phi _ { 1 }$, $\phi _ { 2 }$, and $\phi _ { 3}$ in \eqref{6.17}-\eqref{6.18-1}, we reach that
\begin{align}\label{6.20}
&\rho _ { 1 } \gamma _ { 1 } \int _ { \mathbb { R } } \int _ { - l } ^ { \eta } y \widetilde { \varphi } _ { 1 , x } d y d x + \rho _ { 2 } \gamma _ { 2 } \int _ { \mathbb { R } } \int _ { \eta } ^ { h _ { 2 } +\eta_2 } y \widetilde { \varphi } _ { 2 , x } d y d x \nonumber\\
&= - \int _ { \mathbb { R } } \left( \rho _ { 1 } \gamma _ { 1 } \phi _ { 1 } \eta \eta _ { x } - \rho _ { 2 } \gamma _ { 2 } \phi _ { 2 } \eta \eta _ { x } \right) d x-\gamma _ { 2 } \int _ { \mathbb { R } } \xi_2 (\eta_2+h_2) \eta _ {2, x }dx.
\end{align}
Similarly, the sixth and seventh terms of \eqref{6.4} are reformed as
\begin{equation}\label{6.21}
\rho _ { 1 } \kappa \int _ { \mathbb { R } } \int _ { - l  } ^ { \eta } \tilde { \varphi } _ { 1 , x } d y d x + \rho _ { 2 } \kappa \int _ { \mathbb { R } }\int _ { \eta } ^ { h _ { 2 }+\eta_2 } \tilde { \varphi } _ { 2 , x } d y d x
=-\kappa \int _ { \mathbb { R } } \left( \xi \eta _ { x } +\xi_2\eta_{2,x}\right) d x.
\end{equation}
Summing up \eqref{6.19}, \eqref{6.20} with \eqref{6.21} gives that
\begin{align}\label{6.22}
&\frac{1}{2} \int_{\mathbb{R}} \begin{pmatrix}\xi \\ \xi_{2}\end{pmatrix}^{T}\begin{pmatrix}-G_{11} & -G_{12} \\ G_{21} & G_{22}\end{pmatrix}\begin{pmatrix}\phi_{2} \\ \phi_{3}\end{pmatrix} dx
-\frac{1}{2}\int_{\mathbb{R}} \left(\rho_1 \gamma_1 \phi_1-\rho_{2} \gamma_{2} \phi_{2}\right) \eta \eta_{x} d x\nonumber\\
&-\frac{1}{2}\int_{\mathbb{R}} (\gamma_{2} \eta+2\kappa) \eta_{x}\xi d x
- \int _ { \mathbb { R } } \left( \gamma _ { 2 }(\eta_2+h_2) +\kappa\right)\xi_2\eta _ {2, x }dx\nonumber\\
&=\frac{1}{2} \int_{\mathbb{R}}\begin{pmatrix}\xi \\ \xi_{2}\end{pmatrix}^{T}
\begin{pmatrix}G_{11} \mathcal{B}^{-1} G_1 & \frac{\rho_1}{\rho_{2}} G_{11} \mathcal{B}^{-1} G_{12}-\frac{1}{\rho_{2}} G_{12} \\
-G_{21} \mathcal{B}^{-1} G_1 & -\frac{\rho_1}{\rho_{2}}G_{21} \mathcal{B}^{-1} G_{12}+\frac{1}{\rho_{2}} G_{22}\end{pmatrix}\begin{pmatrix}\xi \\ \xi_{2}\end{pmatrix} d x
\nonumber\\
&+\frac{1}{2} \int_{\mathbb{R}} \begin{pmatrix}\xi \\ \xi_{2}\end{pmatrix}^{T}\begin{pmatrix}-\rho_1\left(\gamma_1-\gamma_{2}\right) G_{11} \mathcal{B}^{-1}\left(\eta \eta_{x}\right) \\ \rho_1\left(\gamma_1-\gamma_{2}\right) G_{21} \mathcal{B}^{-1}\left(\eta \eta_{x}\right)\end{pmatrix} d x -\frac{1}{2}\int_{\mathbb{R}} (\gamma_{2} \eta+2\kappa) \eta_{x}\xi d x\nonumber\\
&- \int _ { \mathbb { R } } \left( \gamma _ { 2 }(\eta_2+h_2) +\kappa\right)\xi_2\eta _ {2, x }dx-\frac{1}{2}\int_{\mathbb{R}} \left(\rho_1( \gamma_1 -\gamma_{2} )\phi_{1}+\gamma_{2}\xi\right) \eta \eta_{x} d x\nonumber\\
&=\frac{1}{2} \int_{\mathbb{R}}\begin{pmatrix}\xi \\ \xi_{2}\end{pmatrix}^{T}\begin{pmatrix}G_{11} \mathcal{B}^{-1} G_1 & -G_1 \mathcal{B}^{-1} G_{12}\\
-G_{21} \mathcal{B}^{-1} G_1 & -\frac{\rho_1}{\rho_{2}}G_{21} \mathcal{B}^{-1} G_{12}+\frac{1}{\rho_{2}} G_{22}\end{pmatrix}\begin{pmatrix}\xi \\ \xi_{2}\end{pmatrix} d x
\nonumber\\
&+\frac{1}{2} \int_{\mathbb{R}} \begin{pmatrix}\xi \\ \xi_{2}\end{pmatrix}^{T}\begin{pmatrix}-\rho_1\left(\gamma_1-\gamma_{2}\right) G_{11} \mathcal{B}^{-1}\left(\eta \eta_{x}\right) \\ \rho_1\left(\gamma_1-\gamma_{2}\right) G_{21} \mathcal{B}^{-1}\left(\eta \eta_{x}\right)\end{pmatrix} d x -\int_{\mathbb{R}} (\gamma_{2} \eta+\kappa) \eta_{x}\xi d x\nonumber\\
&-\frac{1}{2} \int_{\mathbb{R}} \rho_1\left(\gamma_1-\gamma_{2}\right) \eta \eta_{x} \mathcal{B}^{-1}\left(G_{11} \xi-G_{12} \xi_{2}+\rho_{2}\left(\gamma_1-\gamma_{2}\right) \eta\eta_{x}\right) d x \nonumber\\
&- \int _ { \mathbb { R } } \left( \gamma _ { 2 }(\eta_2+h_2) +\kappa\right)\xi_2\eta _ {2, x }dx\nonumber\\
&=\frac{1}{2} \int_{\mathbb{R}}\begin{pmatrix}\xi \\ \xi_{2}\end{pmatrix}^{T}\begin{pmatrix}G_{11} \mathcal{B}^{-1} G_1 & -G_1 \mathcal{B}^{-1} G_{12} \\
-G_{21} \mathcal{B}^{-1} G_1 & -\frac{\rho_1}{\rho_{2}}G_{21} \mathcal{B}^{-1} G_{12}+\frac{1}{\rho_{2}} G_{22}\end{pmatrix}\begin{pmatrix}\xi \\ \xi_{2}\end{pmatrix} d x\\
&+\left(\gamma_1-\gamma_{2}\right)\int _ { \mathbb { R } } \eta \eta_{x} \mathcal{B}^{-1}\left(\rho_{2} G_1 \xi+\rho_1 G_{12} \xi_{2}\right) d x-\frac{\rho_1 \rho_{2} \left(\gamma_1-\gamma_{2}\right)^2}{2} \int _ { \mathbb { R } }  \eta \eta_{x} \mathcal{B}^{-1} \eta \eta_{x}d x\nonumber\\
&- \int _ { \mathbb { R } } \left( \gamma _ { 2 }(\eta_2+h_2) +\kappa\right)\xi_2\eta _ {2, x }dx-\int_{\mathbb{R}} (\gamma_{1} \eta+\kappa) \eta_{x}\xi d x,\nonumber
\end{align}
where the relation
\[
\frac{\rho_1}{\rho_{2}} G_{11} \mathcal{B}^{-1} G_{12}-\frac{1}{\rho_{2}} G_{12}=\frac{1}{\rho_{2}}\left(\mathcal{B}-\rho_{2} G_1\right) \mathcal{B}^{-1} G_{12}-\frac{1}{\rho_{2}} G_{12}=-G_1 \mathcal{B}^{-1} G_{12}
\]
is used and the last equality is obtained by employing that the operators $ B^{-1}, G_{11} $ are self-adjoint, while $ G_{12}^{*}=G_{21} $, cf. \cite{CGK}.

Next, the eighth term of \eqref{6.4} resolves to constants which are added to
the constant Hamiltonian density $\mathfrak{h}_{0}$. The remaining terms of \eqref{6.4} are reformed as
\begin{align}\label{6.23}
&\frac { 1 } { 2 } \rho _ { 1 } \int _ { \mathbb { R } } \int _ { - l } ^ { \eta } \left( \gamma _ { 1 } y + \kappa \right) ^ { 2 } d y d x + \frac { 1 } { 2 } \rho _ { 2 } \int _ { \mathbb { R } } \int _ { \eta } ^ { h _ { 2 } +\eta_2} \left( \gamma _ { 2 } y + \kappa \right) ^ { 2 } d y d x \nonumber\\
&\quad+ \rho _ { 1 } g \int _ { \mathbb { R } } \int _ { B } ^ { \eta } y d y d x + \rho _ { 2 } g \int _ { \mathbb { R } } \int _ { \eta } ^ { h _ { 2 } +\eta_2} y d y dx\nonumber\\
&= \frac { \rho _ { 1 } } { 6 \gamma _ { 1 } } \int _ { \mathbb { R } } \left( \gamma _ { 1 } \eta + \kappa \right) ^ { 3 } d x + \frac { \rho _ { 2 } } { 6 \gamma _ { 2 } } \int _ { \mathbb { R } }\left( \left( \gamma _ { 2 } (\eta_2+h_2) + \kappa \right) ^ { 3 }-\left( \gamma _ { 2 } \eta + \kappa \right) ^ { 3 }\right) d x \nonumber\\
&\quad+ \frac { 1 } { 2 } g \left( \rho _ { 1 } - \rho _ { 2 } \right) \int _ { \mathbb { R } } \eta ^ { 2 } d x-\frac { g \rho _ { 1 } } { 2 } \int _ { \mathbb { R } }B^2(x)dx
+\frac { g \rho _ { 2 } } { 2 } \int _ { \mathbb { R } }(h_2+\eta_2)^2dx\nonumber\\
&= \frac { \rho _ { 2 } \gamma _ { 2}^2 } { 6 } \int _ { \mathbb { R } } \eta_2 ^ { 3 } d x
+\frac {  \rho _ { 1 } \gamma _ { 1}^2-\rho _ { 2 } \gamma _ { 2}^2 } { 6 } \int _ { \mathbb { R } } \eta ^ { 3 } d x
+ \frac { g \left( \rho _ { 1 } - \rho _ { 2 } \right)+\kappa\left( \rho _ { 1 } \gamma _ { 1}- \rho _ { 2 } \gamma _ { 2}\right) } { 2 }  \int _ { \mathbb { R } } \eta ^ { 2 } d x\nonumber\\
&+\frac { g\rho _ { 2 }+\rho _ { 2 } \gamma _ { 2 } ( \gamma _ { 2 } h_2+\kappa) } { 2} \int _ { \mathbb { R } } \eta _ { 2 } ^ { 2 } d x
-\frac { g \rho _ { 1 } } { 2 } \int _ { \mathbb { R } }B^2(x)dx,
\end{align}
where the constants are added to the constant Hamiltonian density $\mathfrak{h}_{0}$.
As the term $\int_{\mathbb{R}} B^{2}(x) \mathrm{d} x$ is a constant and will not contribute to $\delta H$, we obtain from \eqref{6.22}, \eqref{6.23} and \eqref{6.4} that the Hamiltonian $H (\eta,\eta_2,\xi,\xi_2,B(x))$ of the system is
\begin{align}\label{6.24}
&H=\frac{1}{2} \int_{\mathbb{R}}[\frac 1{\rho_2}\xi_{2} G_{22} \xi_{2}+\xi G_1 \mathcal{B}^{-1} G_{11} \xi-\xi G_1 \mathcal{B}^{-1} G_{12} \xi_{2}
-\xi_{2} G_{21} \mathcal{B}^{-1} G_1 \xi \nonumber \\
&-\frac{\rho_1}{\rho_2} \xi_{2} G_{21} \mathcal{B}^{-1} G_{12} \xi_{2}] d x
+ \frac { g \left( \rho _ { 1 } - \rho _ { 2 } \right)+\kappa\left( \rho _ { 1 } \gamma _ { 1}- \rho _ { 2 } \gamma _ { 2}\right) } { 2 }  \int _ { \mathbb { R } } \eta ^ { 2 } d x\nonumber \\
&+\frac { g\rho _ { 2 }+\rho _ { 2 } \gamma _ { 2 } ( \gamma _ { 2 } h_2+\kappa) } { 2} \int _ { \mathbb { R } } \eta _ { 2 } ^ { 2 } d x
+\frac { \rho _ { 2 } \gamma _ { 2}^2 } { 6 } \int _ { \mathbb { R } } \eta_2 ^ { 3 } d x+\frac {  \rho _ { 1 } \gamma _ { 1}^2-\rho _ { 2 } \gamma _ { 2}^2 } { 6 } \int _ { \mathbb { R } } \eta ^ { 3 } d x\nonumber\\
&- \int _ { \mathbb { R } } \left( \gamma _ { 2 }(\eta_2+h_2) +\kappa\right)\xi_2\eta _ {2, x }dx-\int_{\mathbb{R}} (\gamma_{1} \eta+\kappa) \eta_{x}\xi d x \nonumber\\
&+\left(\gamma_1-\gamma_{2}\right)\int _ { \mathbb { R } } \eta \eta_{x} \mathcal{B}^{-1}\left(\rho_{2} G_1 \xi+\rho_1 G_{12} \xi_{2}\right) d x\nonumber\\
&-\frac{\rho_1 \rho_{2} \left(\gamma_1-\gamma_{2}\right)^2}{2} \int _ { \mathbb { R } }  \eta \eta_{x} \mathcal{B}^{-1} \eta \eta_{x}d x.
\end{align}
\subsection{The nearly-Hamiltonian formulation}
In this subsection, we evaluate the variation of the Hamiltonian formulation \eqref{6.4}.
The variation of the first two terms of \eqref{6.4} is
\begin{align}\label{6.25}
&\delta \left[ \frac { 1 } { 2 } \rho _ { 1 } \int _ { \mathbb { R } } \int _ { B } ^ { \eta } \left| \nabla \widetilde { \varphi } _ { 1 } \right| ^ { 2 } d y d x + \frac { 1 } { 2 } \rho _ { 2 } \int _ { \mathbb { R } }  \int _ { \eta } ^ { h _ { 2 } +\eta _ { 2 }} \left| \nabla \tilde { \varphi } _ { 2 } \right| ^ { 2 } d y d x \right] \nonumber\\
&= \rho _ { 1 } \int _ { \mathbb { R } } \int _ { B } ^ { \eta } \left( \nabla \widetilde { \varphi } _ { 1 } \right) \cdot \nabla \delta \widetilde { \varphi } _ { 1 } d y d x
+ \rho _ { 2 } \int _ { \mathbb { R } } \int _ { \eta } ^ { h _ { 2 } +\eta _ { 2 }} \left( \nabla \widetilde { \varphi } _ { 2 } \right) \cdot \nabla \delta \widetilde { \varphi } _ { 2 } d y d x \nonumber\\
&+ \frac { 1 } { 2 } \rho _ { 1 } \int _ { \mathbb { R } } \left| \nabla \widetilde { \varphi } _ { 1 } \right| _ { c } ^ { 2 } \delta \eta d x - \frac { 1 } { 2 } \rho _ { 2 } \int _ { \mathbb { R } } \left| \nabla \widetilde { \varphi } _ { 2 } \right| _ { c } ^ { 2 } \delta \eta d x+ \frac { 1 } { 2 } \rho _ { 2 } \int _ { \mathbb { R } } \left| \nabla \widetilde { \varphi } _ { 2 } \right| _ { s } ^ { 2 } \delta (\eta_2) d x,
\end{align}
where $\delta B=0$ as $B$ is non-dynamic. Note that
\begin{equation*}
\nabla \cdot \left( \left( \nabla \widetilde { \varphi } _ { i } \right) \delta \tilde { \varphi } _ { i } \right)
=\nabla \cdot \left( \nabla \widetilde { \varphi } _ { i } \right) \delta \widetilde { \varphi } _ { i } + \left( \nabla \widetilde { \varphi } _ { i } \right) \cdot \nabla \left( \delta \widetilde { \varphi } _ { i } \right)
= \left( \nabla \widetilde { \varphi } _ { i } \right) \cdot \nabla \left( \delta \widetilde { \varphi } _ { i } \right),
\end{equation*}
where the assumption of incompressibility $\Delta\widetilde { \varphi } _ { i }=0$ is used. Using the divergence theorem
\begin{equation*}
\iint _ { \Omega _ { i } } \nabla \cdot \left( \left( \nabla \widetilde { \varphi } _ { i } \right) \delta \widetilde { \varphi } _ { i } \right) d y d x = \int _ { \mathbb { R } } \left( \left( \nabla \widetilde { \varphi } _ { i } \right) \delta \widetilde { \varphi } _ { i } \right) \cdot \mathbf { n } _ { i } d S,
\end{equation*}
where $\mathbf { n } _ { i }$ for $i=1,2,3$ are the outward normal vector and $dS$ is an infinitesimal surface area, we get
\begin{align}
&\int _ { \mathbb { R } }\int _ { B } ^ { \eta } \left( \nabla \widetilde { \varphi } _ { 1 } \right) \cdot \nabla \delta \widetilde { \varphi } _ { 1 } d y d x\nonumber\\
&=\int _ { \mathbb { R } }\left[ \left[ \begin{array} { l } \left( \widetilde { \varphi } _ { 1 , x } \right) _ { c } \\ \left( \widetilde { \varphi } _ { 1 , y } \right) _ { c } \end{array} \right] \left( \delta \widetilde { \varphi } _ { 1 } \right) _ { c } \right] \cdot \left[ \begin{array} { c } - \eta _ { x } \\ 1 \end{array} \right] d x
+ \int _ { \mathbb { R } } \left[ \left[ \begin{array} { l } \left( \widetilde { \varphi } _ { 1 , x } \right) _ { b } \\ \left( \widetilde { \varphi } _ { 1 , y } \right) _ { b } \end{array} \right] \left( \delta \widetilde { \varphi } _ { 1 } \right) _ { b } \right] \cdot \left[ \begin{array} { c } B'(x) \\ - 1 \end{array} \right] d x,\nonumber
\end{align}
and
\begin{align}
&\int _ { \mathbb { R } }\int _ { \eta } ^ { h_2 +\eta_2} \left( \nabla \widetilde { \varphi } _ { 2 } \right) \cdot \nabla \delta \widetilde { \varphi } _ { 2 } d y d x\nonumber\\
&=\int _ { \mathbb { R } }\left[ \left[ \begin{array} { l } \left( \widetilde { \varphi } _ { 2 , x } \right) _ { c } \\ \left( \widetilde { \varphi } _ { 2 , y } \right) _ { c } \end{array} \right] \left( \delta \widetilde { \varphi } _ { 2 } \right) _ { c } \right] \cdot \left[ \begin{array} { c } \eta _ { x } \\ -1 \end{array} \right] d x + \int _ { \mathbb { R } } \left[ \left[ \begin{array} { l } \left( \widetilde { \varphi } _ { 2 , x } \right) _ { s } \\ \left( \widetilde { \varphi } _ { 2 , y } \right) _ { s } \end{array} \right] \left( \delta \widetilde { \varphi } _ { 2 } \right) _ { s } \right] \cdot \left[ \begin{array} { c } -\eta _ {2, x }  \\  1 \end{array} \right] d x.\nonumber
\end{align}
Using the kinematic boundary conditions on the bed \eqref{2.14}, we rewrite \eqref{6.25} as
\begin{align}\label{6.26}
&\delta \left[ \frac { 1 } { 2 } \rho _ { 1 } \int _ { \mathbb { R } } \int _ { - h _ { 1 } } ^ { \eta } \left| \nabla \widetilde { \varphi } _ { 1 } \right| ^ { 2 } d y d x + \frac { 1 } { 2 } \rho _ { 2 } \int _ { \mathbb { R } } ^ { h _ { 2 } } \int _ { \eta } ^ { \eta } \left| \nabla \widetilde { \varphi } _ { 2 } \right| ^ { 2 } d y d x \right]\nonumber\\
&=\rho _ { 1 } \int _ { \mathbb { R } } \left( \left( \widetilde { \varphi } _ { 1 , y } \right) _ { c } - \left( \widetilde { \varphi } _ { 1 , x } \right) _ { c } \eta _ { x } \right) \left( \delta \widetilde { \varphi } _ { 1 } \right) _ { c } d x - \rho _ { 2 } \int _ { \mathbb { R } } \left( \left( \widetilde { \varphi } _ { 2 , y } \right) _ { c } - \left( \widetilde { \varphi } _ { 2 , x } \right) _ { c } \eta _ { x } \right) \left( \delta \widetilde { \varphi } _ { 2 } \right) _ { c } d x\nonumber\\
&+\rho _ { 2 } \int _ { \mathbb { R } } \left( \left( \widetilde { \varphi } _ { 2 , y } \right) _ { s } - \left( \widetilde { \varphi } _ { 2 , x } \right) _ { s } \eta _ { 2, x } \right) \left( \delta \widetilde { \varphi } _ { 2 } \right) _ { s } d x
+ \frac { 1 } { 2 } \rho _ { 1 } \int _ { \mathbb { R } } \left| \nabla \widetilde { \varphi } _ { 1 } \right| _ { c } ^ { 2 } \delta \eta d x \nonumber\\
&+ \frac { 1 } { 2 } \rho _ { 2 } \int _ { \mathbb { R } } \left| \nabla \widetilde { \varphi } _ { 2 } \right| _ { s } ^ { 2 } \delta (\eta_2) d x- \frac { 1 } { 2 } \rho _ { 2 } \int _ { \mathbb { R } } \left| \nabla \widetilde { \varphi } _ { 2 } \right| _ { c } ^ { 2 } \delta \eta d x.
\end{align}
The variation in the third term of \eqref{6.4} is
\begin{equation*}
\delta  \left[\rho _ { 1 } \int _ { \mathbb { R } } \int _ { - m }  ^ { - l } U \tilde { \varphi } _ { 1 , x } d y d x\right]
= \delta \rho _ { 1 } \int _ { - m } ^ { - l } U _ { 1 } ( y ) \int _ { \mathbb { R } } \tilde { \varphi } _ { 1 , x } d x d y =0.
\end{equation*}
The variation in the fourth and fifth terms of \eqref{6.4} is
\begin{align}\label{6.27}
&\delta \left[ \rho _ { 1 } \gamma _ { 1 }\int _  {\mathbb { R }  } \int _ { -l } ^{ \eta } y \tilde { \varphi } _ { 1 , x } d y d x + \rho _ { 2 } \gamma _ { 2 } \int _  {\mathbb { R }  } \int _ { \eta } ^ { h _ { 2 }+\eta_2 }y \tilde { \varphi } _ { 2 , x } d y d x \right]\nonumber\\
&= \rho _ { 1 } \gamma _ { 1 } \int _ { \mathbb { R } } \eta \left( \widetilde { \varphi } _ { 1 , x } \right) _ { c } \delta \eta d x- \rho _ { 2 } \gamma _ { 2 } \int _ { \mathbb { R } } \eta \left( \tilde { \varphi } _ { 2 , x } \right) _ { c } \delta \eta d x + \rho _ { 2 } \gamma _ { 2 } \int _ { \mathbb { R } } (h_2+\eta_2) \left( \tilde { \varphi } _ { 2 , x } \right) _ { s } \delta (\eta_2) d x \nonumber\\
&\quad+ \rho _ { 1 } \gamma _ { 1 } \int _ { \mathbb { R } } \int _ { - l } ^ { \eta } y \delta \left( \widetilde { \varphi } _ { 1 , x } \right) d y d x + \rho _ { 2 } \gamma _ { 2 } \int _ { \mathbb { R } } \int _ { \eta } ^ { h _ { 2 }+\eta_2 } y \delta \left( \tilde { \varphi } _ { 2 , x } \right) d y d x\nonumber\\
&= \rho _ { 1 } \gamma _ { 1 } \int _ { \mathbb { R } } \eta \left( \widetilde { \varphi } _ { 1 , x } \right) _ { c } \delta \eta d x- \rho _ { 2 } \gamma _ { 2 } \int _ { \mathbb { R } } \eta \left( \tilde { \varphi } _ { 2 , x } \right) _ { c } \delta \eta d x + \rho _ { 2 } \gamma _ { 2 } \int _ { \mathbb { R } } (h_2+\eta_2) \left( \tilde { \varphi } _ { 2 , x } \right) _ { s } \delta (\eta_2) d x\nonumber\\
&\quad- \rho _ { 1 } \gamma _ { 1 } \int _ { \mathbb { R } } \eta \eta _ { x } \left( \delta \widetilde { \varphi } _ { 1 } \right) _ { c } d x + \rho _ { 2 } \gamma _ { 2 } \int _ { \mathbb { R } } \eta \eta _ { x } \left( \delta \widetilde { \varphi } _ { 2 } \right) _ { c } d x\nonumber\\
&- \rho _ { 2 } \gamma _ { 2 } \int _ { \mathbb { R } } (h_2+\eta_2) \eta _ { 2, x } \left( \delta \widetilde { \varphi } _ { 2 } \right) _ { s } d x.
\end{align}
The variation in the sixth and seventh terms of \eqref{6.4} is
\begin{align}\label{6.28}
&\delta \left[ \rho _ { 1 } \kappa \int _ { \mathbb { R } } \int _ { - l } ^ { \eta } \tilde { \varphi } _ { 1 , x } d y d x + \rho _ { 2 } \kappa \int _ { \mathbb { R } } \int _ { \eta } ^ { h _ { 2 }+\eta_2 } \tilde { \varphi } _ { 2 , x } d y d x \right]\nonumber \\
&= \rho _ { 1 } \kappa \int _ { \mathbb { R } } \left( \widetilde { \varphi } _ { 1 , x } \right) _ { c } \delta \eta d x- \rho _ { 2 } \kappa \int _ { \mathbb { R } } \left( \widetilde { \varphi } _ { 2 , x } \right) _ { c } \delta \eta d x
+\rho _ { 2 } \kappa \int _ { \mathbb { R } } \left( \widetilde { \varphi } _ { 2 , x } \right) _ {s } \delta (\eta_2) d x \nonumber\\
&\quad+ \rho _ { 1 } \kappa \int _ { \mathbb { R } } \int _ { - l } ^ { \eta } \delta \left( \widetilde { \varphi } _ { 1 , x } \right) d y d x + \rho _ { 2 } \kappa\int _ { \mathbb { R } } \int _ { \eta } ^ { h _ { 2 }+\eta_2} \delta \left( \widetilde { \varphi } _ { 2 , x } \right) d y d x\nonumber\\
&=\rho _ { 1 } \kappa\int _ { \mathbb { R } } \left( \widetilde { \varphi } _ { 1 , x } \right) _ { c } \delta \eta d x- \rho _ { 2 } \kappa \int _ { \mathbb { R } } \left( \widetilde { \varphi } _ { 2 , x } \right) _ { c } \delta \eta d x
+\rho _ { 2 } \kappa \int _ { \mathbb { R } } \left( \widetilde { \varphi } _ { 2 , x } \right) _ { s } \delta (\eta_2) d x\nonumber\\
&\quad- \rho _ { 1 } \kappa \int _ { \mathbb { R } } \eta _ { x } \left( \delta \tilde { \varphi } _ { 1 } \right) _ { c } d x + \rho _ { 2 } \kappa \int _ { \mathbb { R } } \eta _ { x } \left( \delta \widetilde { \varphi } _ { 2 } \right) _ { c } dx
- \rho _ { 2 } \kappa \int _ { \mathbb { R } } \eta _ {2, x } \left( \delta \widetilde { \varphi } _ { 2 } \right) _ { s } d x.
\end{align}
The variation in the eighth term of \eqref{6.4} is
\begin{equation*}
\delta \left[\frac { 1 } { 2 } \rho _ { 1 } \int _ { \mathbb { R } } \int _ { - m _ { 1 } } ^ { - l _ { 1 } } U _ { 1 } ^ { 2 } d y d x \right] = Const.,
\end{equation*}
which will not contribute to the dynamics. The variations in the remaining terms of \eqref{6.4} are
\begin{align}\label{6.29}
&\delta \left[ \frac { 1 } { 2 } \rho _ { 1 } \gamma _ { 1 } ^ { 2 } \int _ { \mathbb { R } } \int _ { - l } ^ { \eta } y ^ { 2 } d y d x + \frac { 1 } { 2 } \rho _ { 2 } \gamma _ { 2 } ^ { 2 } \int _ { \mathbb { R } }  \int _ { \eta }  ^ { h _ { 2 }+\eta_2 }y^2d y d x \right]\nonumber\\
&= \frac { 1 } { 2 } \rho _ { 1 } \gamma _ { 1 } ^ { 2 } \int _ { \mathbb { R } } \eta ^ { 2 } \delta \eta d x - \frac { 1 } { 2 } \rho _ { 2 } \gamma _ { 2 } ^ { 2 } \int _ { \mathbb { R } } \eta ^ { 2 } \delta \eta d x
+\frac { 1 } { 2 } \rho _ { 2 } \gamma _ { 2 } ^ { 2 } \int _ { \mathbb { R } } (h _ { 2 }+\eta_2 ) ^ { 2 } \delta (\eta_2) d x,\nonumber\\
&\delta \left[ \frac { 1 } { 2 } \rho _ { 1 } \kappa ^ { 2 } \int _ { \mathbb { R } } \int _ { - l } ^ { \eta } d y d x + \frac { 1 } { 2 } \rho _ { 2 } \kappa  ^ { 2 } \int _ { \mathbb { R } }  \int _ { \eta }  ^ { h _ { 2 }+\eta_2 } d y d x \right] \nonumber\\
&= \frac { 1 } { 2 } \rho _ { 1 } \kappa  ^ { 2 } \int _ { \mathbb { R } } \delta \eta d x + \frac { 1 } { 2 } \rho _ { 2 } \kappa  ^ { 2 } \int _ { \mathbb { R } } \delta (\eta_2- \eta) d x,\nonumber\\
&\delta \left[ \rho _ { 1 } \gamma _ { 1 } \kappa \int _ { \mathbb { R } }  \int _ { B }  ^ { \eta }y d x + \rho _ { 2 } \gamma _ { 2 } \kappa \int _ { \mathbb { R } } \int _ { \eta }  ^ { h _ { 2 }+\eta_2 } y d x + \rho _ { 1 } g \int _ { \mathbb { R } }  \int _ { B }  ^ { \eta } y d x + \rho _ { 2 } g \int _ { \mathbb { R } } \int _ { \eta }  ^ { h _ { 2 }+\eta_2 } y d x \right]\nonumber\\
&= \rho _ { 1 } \left( g + \gamma _ { 1 } \kappa \right) \int _ { \mathbb { R } } \eta \delta \eta d x+ \rho _ { 2 } \left( g + \gamma _ { 2 } \kappa  \right) \int _ { \mathbb { R } } \left((h_2+\eta_2) \delta (\eta_2)-\eta \delta \eta \right) d x,
\end{align}
plus some constants which will not contribute to the dynamics. Employing \eqref{6.26}-\eqref{6.29}, we get the variation of the Hamiltonian as
\begin{align}\label{6.30}
&\delta H = \int _ { \mathbb { R } } \Big( \rho _ { 1 } \left( \left( \widetilde { \varphi } _ { 1 , y } \right) _ { c } - \left( \widetilde { \varphi } _ { 1 , x } \right) _ { c } \eta _ { x } \right) \left( \delta \widetilde { \varphi } _ { 1 } \right) _ { c } - \rho _ { 2 } \left( \left( \widetilde { \varphi } _ { 2 , y } \right) _ { c } - \left( \widetilde { \varphi } _ { 2 , x } \right) _ { c } \eta _ { x } \right) \left( \delta \widetilde { \varphi } _ { 2 } \right) _ { c } \nonumber\\
&+\rho _ { 2 } \left( \left( \widetilde { \varphi } _ { 2 , y } \right) _ { s } - \left( \widetilde { \varphi } _ { 2 , x } \right) _ { s } \eta _ { 2, x } \right) \left( \delta \widetilde { \varphi } _ { 2 } \right) _ { s } + \frac { 1 } { 2 } \rho _ { 2 } \left| \nabla \widetilde { \varphi } _ { 2 } \right| _ { s } ^ { 2 } \delta (\eta_2)  +\rho _ { 2 } \kappa  \left( \widetilde { \varphi } _ { 2 , x } \right) _ { s } \delta (\eta_2)\nonumber \\
&- \rho _ { 2 } \kappa \eta _ {2, x } \left( \delta \widetilde { \varphi } _ { 2 } \right) _ { s } + \rho _ { 2 } \gamma _ { 2 } (h_2+\eta_2) \left( \tilde { \varphi } _ { 2 , x } \right) _ { s } \delta (\eta_2) - \rho _ { 2 } \gamma _ { 2 }(h_2+\eta_2) \eta _ { 2, x } \left( \delta \widetilde { \varphi } _ { 2 } \right) _ { s }\nonumber\\
&+\frac { 1 } { 2 } \rho _ { 2 } \gamma _ { 2 } ^ { 2 } (h _ { 2 }+\eta_2 ) ^ { 2 } \delta (\eta_2)
+ \frac { 1 } { 2 } \rho _ { 2 } \kappa  ^ { 2 }  \delta (\eta_2)
+ \rho _ { 2 } \left( g + \gamma _ { 2 } \kappa \right) (h_2+\eta_2) \delta (\eta_2)
\nonumber\\
&+ \frac { 1 } { 2 } \rho _ { 1 } \left| \nabla \widetilde { \varphi } _ { 1 } \right| _ { c } ^ { 2 } \delta \eta - \frac { 1 } { 2 } \rho _ { 2 } \left| \nabla \widetilde { \varphi } _ { 2 } \right| _ { c } ^ { 2 } \delta \eta+ \rho _ { 1 } \gamma _ { 1 } \eta \left( \widetilde { \varphi } _ { 1 , x } \right) _ { c } \delta \eta- \rho _ { 2 } \gamma _ { 2 } \eta \left( \widetilde { \varphi } _ { 2 , x } \right) _ { c } \delta \eta\nonumber\\
&- \rho _ { 1 } \gamma _ { 1 } \eta \eta _ { x } \left( \delta \widetilde { \varphi } _ { 1 } \right) _ { c }+ \rho _ { 2 } \gamma _ { 2 }  \eta \eta _ { x } \left( \delta \widetilde { \varphi } _ { 2 } \right) _ { c }+ \rho _ { 1 } \kappa \left( \widetilde { \varphi } _ { 1 , x } \right) _ { c } \delta \eta - \rho _ { 2 } \kappa  \left( \widetilde { \varphi } _ { 2 , x } \right) _ { c } \delta \eta \nonumber\\
&- \rho _ { 1 } \kappa  \eta _ { x } \left( \delta \tilde { \varphi } _ { 1 } \right) _ { c } + \rho _ { 2 } \kappa \eta _ { x } \left( \delta \widetilde { \varphi } _ { 2 } \right) _ { c }+\frac { 1 } { 2 } \rho _ { 1 } \gamma _ { 1 } ^ { 2 }  \eta ^ { 2 } \delta \eta- \frac { 1 } { 2 } \rho _ { 2 } \gamma _ { 2 } ^ { 2 }  \eta ^ { 2 } \delta \eta\nonumber\\
&+\frac { 1 } { 2 } \rho _ { 1 } \kappa ^ { 2 }  \delta \eta - \frac { 1 } { 2 } \rho _ { 2 } \kappa ^ { 2 }  \delta \eta
+\rho _ { 1 } \left( g + \gamma _ { 1 } \kappa \right)\eta \delta \eta - \rho _ { 2 } \left( g + \gamma _ { 2 } \kappa \right)  \eta \delta \eta \Big)dx\nonumber\\
&= \int _ { \mathbb { R } } \Big( \rho _ { 1 } \eta _ { t } \left( \delta \widetilde { \varphi } _ { 1 } \right) _ { c } +\rho _ { 2 } \eta _ { 2, t } \left( \delta \widetilde { \varphi } _ { 2 } \right) _ { s }- \rho _ { 2 } \eta _ { t } \left( \delta \tilde { \varphi } _ { 2 } \right) _ { c } + \frac { 1 } { 2 } \rho _ { 1 } \left| \nabla \widetilde { \varphi } _ { 1 } \right| _ { c } ^ { 2 } \delta \eta - \frac { 1 } { 2 } \rho _ { 2 } \left| \nabla \widetilde { \varphi } _ { 2 } \right| _ { c } ^ { 2 } \delta \eta\nonumber\\
&+ \frac { 1 } { 2 } \rho _ { 2 } \left| \nabla \widetilde { \varphi } _ { 2 } \right| _ { s } ^ { 2 } \delta (\eta_2)  +\rho _ { 2 } \kappa  \left( \widetilde { \varphi } _ { 2 , x } \right) _ { s } \delta (\eta_2) + \rho _ { 2 } \gamma _ { 2 } (h_2+\eta_2) \left( \tilde { \varphi } _ { 2 , x } \right) _ { s } \delta (\eta_2)\nonumber\\
&+\frac { 1 } { 2 } \rho _ { 2 } \gamma _ { 2 } ^ { 2 } (h _ { 2 }+\eta_2 ) ^ { 2 } \delta (\eta_2)
+ \frac { 1 } { 2 } \rho _ { 2 } \kappa  ^ { 2 }  \delta (\eta_2)
+ \rho _ { 2 } \left( g + \gamma _ { 2 } \kappa \right) (h_2+\eta_2) \delta (\eta_2)
\nonumber\\
&+\rho _ { 1 } \gamma _ { 1 } \eta \left( \widetilde { \varphi } _ { 1 , x } \right) _ { c } \delta \eta - \rho _ { 2 } \gamma _ { 2 } \eta \left( \widetilde { \varphi } _ { 2 , x } \right) _ { c } \delta \eta+ \rho _ { 1 } \kappa \left( \widetilde { \varphi } _ { 1 , x } \right) _ { c } \delta \eta - \rho _ { 2 } \kappa\left( \widetilde { \varphi } _ { 2 , x } \right) _ { c } \delta \eta+\frac { 1 } { 2 } \rho _ { 1 } \gamma _ { 1 } ^ { 2 }  \eta ^ { 2 } \delta \eta\nonumber\\
&- \frac { 1 } { 2 } \rho _ { 2 } \gamma _ { 2 } ^ { 2 }  \eta ^ { 2 } \delta \eta+\frac { 1 } { 2 } \rho _ { 1 } \kappa^ { 2 }  \delta \eta - \frac { 1 } { 2 } \rho _ { 2 } \kappa ^ { 2 }  \delta \eta
+\rho _ { 1 } \left( g + \gamma _ { 1 } \kappa  \right) \eta \delta \eta - \rho _ { 2 } \left( g + \gamma _ { 2 } \kappa \right)  \eta \delta \eta \Big)dx\nonumber\\
&= \int _ { \mathbb { R } } \Big(- \rho _ { 1 } \eta _ { t } \left( \tilde { \varphi } _ { 1 , y } \right) _ { c } + \rho _ { 2 } \eta _ { t } \left( \tilde { \varphi } _ { 2 , y } \right) _ { c }
+ \frac { 1 } { 2 } \rho _ { 1 } \left| \nabla \widetilde { \varphi } _ { 1 } \right| _ { c } ^ { 2 } - \frac { 1 } { 2 } \rho _ { 2 } \left| \nabla \widetilde { \varphi } _ { 2 } \right| _ { c } ^ { 2 }+\rho _ { 1 } \gamma _ { 1 } \eta \left( \widetilde { \varphi } _ { 1 , x } \right) _ { c } \nonumber\\
&- \rho _ { 2 } \gamma _ { 2 } \eta \left( \widetilde { \varphi } _ { 2 , x } \right) _ { c }+ \rho _ { 1 } \kappa \left( \widetilde { \varphi } _ { 1 , x } \right) _ { c } - \rho _ { 2 } \kappa \left( \widetilde { \varphi } _ { 2 , x } \right) _ { c } +\frac { 1 } { 2 } \rho _ { 1 } \gamma _ { 1 } ^ { 2 }  \eta ^ { 2 }- \frac { 1 } { 2 } \rho _ { 2 } \gamma _ { 2 } ^ { 2 }  \eta ^ { 2 }+\frac { 1 } { 2 } \rho _ { 1 } \kappa ^ { 2 }  - \frac { 1 } { 2 } \rho _ { 2 } \kappa^ { 2 }\nonumber\\
&+\rho _ { 1 } \left( g + \gamma _ { 1 } \kappa \right) \eta - \rho _ { 2 } \left( g + \gamma _ { 2 } \kappa \right)  \eta \Big)\delta \eta dx+\int _ { \mathbb { R } } \Big(- \rho _ { 2 } \eta _ { 2, t } \left( \tilde { \varphi } _ { 1 , y } \right) _ { s }+ \frac { 1 } { 2 } \rho _ { 2 } \left| \nabla \widetilde { \varphi } _ { 2 } \right| _ { s } ^ { 2 }\nonumber\\
&+ \rho _ { 2 } \gamma _ { 2 } (h_2+\eta_2) \left( \tilde { \varphi } _ { 2 , x } \right) _ { s }+\rho _ { 2 } \kappa  \left( \widetilde { \varphi } _ { 2 , x } \right) _ { s }
+\frac { 1 } { 2 } \rho _ { 2 } \gamma _ { 2 } ^ { 2 } (h _ { 2 }+\eta_2 ) ^ { 2 } + \frac { 1 } { 2 } \rho _ { 2 } \kappa  ^ { 2 }\nonumber\\
&+ \rho _ { 2 } \left( g + \gamma _ { 2 } \kappa \right) (h_2+\eta_2)\Big)
\delta (\eta_2) dx\nonumber\\
&+ \rho _ { 1 } \int _ { \mathbb { R } } \eta _ { t } \delta \phi _ { 1 } d x - \rho _ { 2 } \int _ { \mathbb { R } } \eta _ { t } \delta \phi _ { 2 } d x
+ \rho _ { 2 } \int _ { \mathbb { R } } \eta _ {2, t } \delta \phi _ { 3 } d x,
\end{align}
where the kinematic boundary conditions \eqref{2.13}, \eqref{2.15}
\begin{align*}
&\rho _ { i } \left( \left( \widetilde { \varphi } _ { i , y } \right) _ { c } - \left( \widetilde { \varphi } _ { i , x } \right) _ { c } \eta _ { x } \right) - \rho _ { i } \gamma _ { i } \eta \eta _ { x } - \rho _ { i } \kappa \eta _ { x } = \rho _ { i } \eta _ { t },\\
&\rho _ { 2}\left(\widetilde{\varphi}_{2, y}\right)_{s}-\rho _ { 2}\left(\left(\widetilde{\varphi}_{2, x}\right)_{s}+\gamma_{2} (h_2+\eta_2)+\kappa\right) \eta_{2,x}=\rho _ { 2}\eta_{2,t},
\end{align*}
and the variation of the velocity potential on the wave
\begin{equation*}
\left( \delta \widetilde { \varphi } _ { i } \right) _ { c } = \delta \phi _ { i } - \left( \widetilde { \varphi } _ { i , y } \right) _ { c } \delta \eta \quad\text{and}\quad
\left( \delta \widetilde { \varphi } _ { 2 } \right) _ { s} = \delta \phi _ { 3 } - \left( \widetilde { \varphi } _ { 2 , y } \right) _ { s } \delta (\eta_2)
\end{equation*}
are used. Using the equalities
\begin{align*}
&\frac { 1 } { 2 } \left| \nabla \widetilde { \varphi } _ { i } \right| _ { c } ^ { 2 } + \frac { 1 } { 2 } \gamma _ { i } ^ { 2 } \eta ^ { 2 } + \gamma _ { i } \eta \left( \widetilde { \varphi } _ { i , x } \right) _ { c } = \frac { 1 } { 2 } \left| \nabla \psi _ { i } \right| _ { c } ^ { 2 } - \frac { 1 } { 2 } \kappa ^ { 2 } - \gamma _ { i } \kappa  \eta - \kappa \left( \widetilde { \varphi } _ { i , x } \right) _ { c },\\
&\frac { 1 } { 2 } \left| \nabla \widetilde { \varphi } _ { 2 } \right| _ { s } ^ { 2 } + \frac { 1 } { 2 } \gamma _ { 2 } ^ { 2 } (h_2+\eta_2) ^ { 2 } + \gamma _ { 2 } (h_2+\eta_2) \left( \widetilde { \varphi } _ { 2 , x } \right) _ { s } \\
&= \frac { 1 } { 2 } \left| \nabla \psi _ { 2 } \right| _ { s } ^ { 2 } - \frac { 1 } { 2 } \kappa ^ { 2 } - \gamma _ { 2 } \kappa  (h_2+\eta_2) - \kappa \left( \widetilde { \varphi } _ { 2 , x } \right) _ { s },
\end{align*}
which are obtained by the definitions of  $\psi_i$ and $\widetilde { \varphi }_i$, we can compute the variation with respect to $\eta$  and $\eta_2$ as
\begin{align}\label{6.31}
\frac { \delta H } { \delta \eta }
&=  - \rho _ { 1 } \eta _ { t } \left( \widetilde { \varphi } _ { 1 , y } \right) _ { c } + \rho _ { 2 } \eta _ { t } \left( \widetilde { \varphi } _ { 2 , y } \right) _ { c } + \frac { 1 } { 2 } \rho _ { 1 } \left| \nabla \widetilde { \varphi } _ { 1 } \right| _ { c } ^ { 2 } - \frac { 1 } { 2 } \rho _ { 2 } \left| \nabla \widetilde { \varphi } _ { 2 } \right| _ { c } ^ { 2 } + \rho _ { 1 } \gamma _ { 1 } \eta \left( \widetilde { \varphi } _ { 1 , x } \right) _ { c }\nonumber\\
&- \rho _ { 2 } \gamma _ { 2 } \eta \left( \widetilde { \varphi } _ { 2 , x } \right) _ { c } + \rho _ { 1 } \kappa \left( \widetilde { \varphi } _ { 1 , x } \right) _ { c } - \rho _ { 2 } \kappa \left( \widetilde { \varphi } _ { 2 , x } \right) _ { c } + \frac { 1 } { 2 } \rho _ { 1 } \gamma _ { 1 } ^ { 2 } \eta ^ { 2 }- \frac { 1 } { 2 } \rho _ { 2 } \gamma _ { 2 } ^ { 2 } \eta ^ { 2 }\nonumber\\
&+ \frac { 1 } { 2 } \rho _ { 1 } \kappa ^ { 2 } - \frac { 1 } { 2 } \rho _ { 2 } \kappa ^ { 2 } + \rho _ { 1 } \left( g + \gamma _ { 1 } \kappa \right) \eta - \rho _ { 2 } \left( g + \gamma _ { 2 } \kappa  \right) \eta\nonumber\\
&= - \rho _ { 1 } \eta _ { t } \left( \widetilde { \varphi } _ { 1 , y } \right) _ { c } + \rho _ { 2 } \eta _ { t } \left( \widetilde { \varphi } _ { 2 , y } \right) _ { c } + \frac { 1 } { 2 } \rho _ { 1 } \left| \nabla \psi _ { 1 } \right| _ { c } ^ { 2 } - \frac { 1 } { 2 } \rho _ { 2 } \left| \nabla \psi _ { 2 } \right| _ { c } ^ { 2 } + \left( \rho _ { 1 } - \rho _ { 2 } \right) g \eta,
\end{align}
and
\begin{align}\label{6.31-1}
\frac { \delta H } { \delta \eta_2 }
&= - \rho _ { 2 } \eta _ { 2, t } \left( \tilde { \varphi } _ { 1 , y } \right) _ { s }+ \frac { 1 } { 2 } \rho _ { 2 } \left| \nabla \widetilde { \varphi } _ { 2 } \right| _ { s } ^ { 2 }
+ \rho _ { 2 } \gamma _ { 2 } (h_2+\eta_2) \left( \tilde { \varphi } _ { 2 , x } \right) _ { s }+\rho _ { 2 } \kappa  \left( \widetilde { \varphi } _ { 2 , x } \right) _ { s }\nonumber\\
&+\frac { 1 } { 2 } \rho _ { 2 } \gamma _ { 2 } ^ { 2 } (h _ { 2 }+\eta_2 ) ^ { 2 } + \frac { 1 } { 2 } \rho _ { 2 } \kappa  ^ { 2 }
+ \rho _ { 2 } \left( g + \gamma _ { 2 } \kappa \right) (h_2+\eta_2)\nonumber\\
&= - \rho _ { 2 } \eta _ { 2, t } \left( \widetilde { \varphi } _ { 2 , y } \right) _ { s }  + \frac { 1 } { 2 } \rho _ { 2 } \left| \nabla \psi _ { 2 } \right| _ { s } ^ { 2 } + \rho _ { 2 } g\left(h_2+\eta_2\right).
\end{align}
From the Bernoulli condition \eqref{2.11}
\begin{equation*}
\begin{aligned} \frac { 1 } { 2 } \rho _ { 1 } \left| \nabla \psi _ { 1 } \right| _ { c } ^ { 2 } - \frac { 1 } { 2 } & \rho _ { 2 } \left| \nabla \psi _ { 2 } \right| _ { c } ^ { 2 } + \left( \rho _ { 1 } - \rho _ { 2 } \right) g \eta \\ & = - \rho _ { 1 } \left( \widetilde { \varphi } _ { 1 , t } \right) _ { c } + \rho _ { 2 } \left( \widetilde { \varphi } _ { 2 , t } \right) _ { c } + \rho _ { 1 } \left( \gamma _ { 1 } + 2 \omega \right) \chi - \rho _ { 2 } \left( \gamma _ { 2 } + 2 \omega \right) \chi,
\end{aligned}
\end{equation*}
we get
\begin{align*}
\frac { \delta H } { \delta \eta } &= - \rho _ { 1 } \eta _ { t } \left( \widetilde { \varphi } _ { 1 , y } \right) _ { c } + \rho _ { 2 } \eta _ { t } \left( \widetilde { \varphi } _ { 2 , y } \right) _ { c } - \rho _ { 1 } \left( \widetilde { \varphi } _ { 1 , t } \right) _ { c } + \rho _ { 2 } \left( \widetilde { \varphi } _ { 2 , t } \right) _ { c } \\
&\quad\;+ \rho _ { 1 } \left( \gamma _ { 1 } + 2 \omega \right) \chi - \rho _ { 2 } \left( \gamma _ { 2 } + 2 \omega \right) \chi,
\end{align*}
and from \eqref{2.11-1} we obtain
\begin{equation*}
\frac { \delta H } { \delta \eta_2 } =- \rho _ { 2 } \eta _ {2, t } \left( \widetilde { \varphi } _ { 2 , y } \right) _ { s }- \rho _ { 2 } \left( \widetilde { \varphi } _ { 2 , t } \right) _ { s }+ \rho _ { 2 } \left( \gamma _ { 2 } + 2 \omega \right) \chi _ { 2 }.
\end{equation*}
Since
\begin{equation*}
\phi _ { i , t } = \left( \widetilde { \varphi } _ { i , t } \right) _ { c } + \left( \widetilde { \varphi } _ { i , y } \right) _ { c } \eta _ { t }, \quad \text{and} \quad
\phi _ { 3 , t } = \left( \widetilde { \varphi } _ { 2 , t } \right) _ { s } + \left( \widetilde { \varphi } _ { 2 , y } \right) _ { s } \eta _ {2,  t },
\end{equation*}
then we get that
\begin{equation}\label{6.33}
\frac { \delta H } { \delta \eta } = - \xi _ { t } + \Gamma \chi,\quad \text{and}\quad
\frac { \delta H } { \delta \eta_2 }  = - \xi _ { 2, t } + \Gamma_2 \chi _ { 2 },
\end{equation}
with $\Gamma$ and $\Gamma_2$ being a new constant defined as
\begin{equation}
\Gamma : = \rho _ { 1 } \gamma _ { 1 } - \rho _ { 2 } \gamma _ { 2 } + 2 \omega \left( \rho _ { 1 } - \rho _ { 2 } \right),\quad \text{and}\quad
\Gamma_2 : = \rho _ { 2 } \left( \gamma _ { 2 } + 2 \omega \right).
\end{equation}
Fixing $\eta$ and $\eta_2$, we obtain from \eqref{6.30} that
\begin{equation}\label{6.34}
\frac { \delta H } { \delta \xi} = \eta _ { t }\quad \text{and} \quad
\frac { \delta H } { \delta \xi_2} = \eta _ {2, t }.
\end{equation}
Equations \eqref{6.33} and \eqref{6.34} therefore give the non-canonical system
\begin{equation}\label{6.35}
\begin{cases}
\eta _ { t } =\frac { \delta H } { \delta \xi},\quad \xi _ { t } = -\frac { \delta H } { \delta \eta } + \Gamma \chi,\\
\eta _ { 2, t } =\frac { \delta H } { \delta \xi_2},\quad \xi _ { 2, t } = -\frac { \delta H } { \delta \eta_2 } + \Gamma_2 \chi_2.
\end{cases}
\end{equation}
\subsection{The Hamiltonian formulation}
Now, we introduce a variable transformation as
\begin{equation}\label{6.36}
\zeta = \xi + \frac { \Gamma } { 2 } \int _ { - \infty } ^ { x } \eta \left( x ^ { \prime } , t \right) d x ^ { \prime}\quad \text{and} \quad \zeta_2 = \xi_2 + \frac { \Gamma_2 } { 2 } \int _ { - \infty } ^ { x } \eta_2 \left( x ^ { \prime } , t \right) d x ^ { \prime}
\end{equation}
to achieve a canonical Hamiltonian system. Indeed,
\begin{align}\label{6.37}
&\delta H \nonumber\\
&= \int _ { \mathbb { R } } \left( - \xi _ { t } + \Gamma \chi \right) \delta \eta d x + \int _ { \mathbb { R } } \eta _ { t } \delta \xi d x+\int _ { \mathbb { R } } \left( - \xi _ { 2, t } + \Gamma_2 \chi_2 \right) \delta \eta_2 d x + \int _ { \mathbb { R } } \eta _ {2, t } \delta \xi_2 d x\nonumber\\
&=\int _ { \mathbb { R } } \left( - \zeta _ { t } + \frac { \Gamma } { 2 } \int _ { - \infty } ^ { x } \eta _ { t } \left( x ^ { \prime } \right) d x ^ { \prime } - \Gamma  \int _ { - \infty } ^ { x } \eta _ { t } ( x ^ { \prime } ) d x ^ { \prime } \right) \delta \eta d x
\nonumber \\
&+ \int _ { \mathbb { R } } \left( \eta _ { t } \delta \zeta - \frac { \Gamma } { 2 } \eta _ { t } \int _ { - \infty } ^ { x } \delta \eta \left( x ^ { \prime } \right) d x ^ { \prime } \right) d x+ \int _ { \mathbb { R } } \left( \eta _ { 2, t } \delta \zeta_2 - \frac { \Gamma_2 } { 2 } \eta _ {2, t } \int _ { - \infty } ^ { x } \delta \eta_2 \left( x ^ { \prime } \right) d x ^ { \prime } \right) d x\nonumber\\
&+\int _ { \mathbb { R } } \left( - \zeta _ { 2, t } + \frac { \Gamma_2 } { 2 } \int _ { - \infty } ^ { x } \eta _ { t } \left( x ^ { \prime } \right) d x ^ { \prime } - \Gamma_2  \int _ { - \infty } ^ { x } \eta _ { 2, t } ( x ^ { \prime } ) d x ^ { \prime } \right) \delta \eta_2 d x \nonumber\\
&=\int _ { \mathbb { R } } \left( - \zeta _ { t } - \frac { \Gamma } { 2 } \int _ { - \infty } ^ { x } \eta _ { t } \left( x ^ { \prime } \right) d x ^ { \prime } \right) \delta \eta d x+ \int _ { \mathbb { R } } \left( \eta _ { t } \delta \zeta - \frac { \Gamma } { 2 } \eta _ { t } \int _ { - \infty } ^ { x } \delta \eta \left( x ^ { \prime } \right) d x ^ { \prime } \right) d x\nonumber\\
&+\int _ { \mathbb { R } } \left( - \zeta _ {2, t } - \frac { \Gamma _2} { 2 } \int _ { - \infty } ^ { x } \eta _ { 2, t } \left( x ^ { \prime } \right) d x ^ { \prime } \right) \delta \eta_2 d x
+ \int _ { \mathbb { R } } \left( \eta _ { 2, t } \delta \zeta_2 - \frac { \Gamma_2 } { 2 } \eta _ { 2, t } \int _ { - \infty } ^ { x } \delta \eta_2 \left( x ^ { \prime } \right) d x ^ { \prime } \right) d x\nonumber\\
&=\int _ { \mathbb { R } } \left( - \zeta _ { t }\delta \eta+\eta _ { t } \delta \zeta\right) d x
+\int _ { \mathbb { R } } \left( - \zeta _ { 2, t }\delta \eta_2+\eta _ { 2, t } \delta \zeta_2\right) d x,
\end{align}
where the integrations by parts
\begin{equation*}
\begin{aligned}
&\int _ { \mathbb { R } } \eta _ { t } \left( \int _ { - \infty } ^ { x } \delta \eta \left( x ^ { \prime } \right) d x ^ { \prime } \right) d x \\
&\quad = \left[ \int _ { - \infty } ^ { x } \delta \eta \left( x ^ { \prime } \right) d x ^ { \prime } \int _ { - \infty } ^ { x } \eta _ { t } \left( x ^ { \prime \prime } \right) d x ^ { \prime \prime } \right] _ { - \infty } ^ { + \infty } - \int _ { \mathbb { R } } \left( \int _ { - \infty } ^ { x } \eta _ { t } \left( x ^ { \prime \prime } \right) d x ^ { \prime \prime } \right) \delta \eta d x
\end{aligned}
\end{equation*}
and
\begin{equation*}
\begin{aligned}
&\int _ { \mathbb { R } } \eta _ {2, t } \left( \int _ { - \infty } ^ { x } \delta \eta_2 \left( x ^ { \prime } \right) d x ^ { \prime } \right) d x \\
&\quad = \left[ \int _ { - \infty } ^ { x } \delta \eta_2 \left( x ^ { \prime } \right) d x ^ { \prime } \int _ { - \infty } ^ { x } \eta _ { 2, t } \left( x ^ { \prime \prime } \right) d x ^ { \prime \prime } \right] _ { - \infty } ^ { + \infty } - \int _ { \mathbb { R } } \left( \int _ { - \infty } ^ { x } \eta _ {2, t } \left( x ^ { \prime \prime } \right) d x ^ { \prime \prime } \right) \delta \eta_2 d x
\end{aligned}
\end{equation*}
are used. And this gives the canonical Hamiltonian system
\begin{equation}\label{6.39}
\begin{cases}
\zeta _ { t } = - \frac { \delta H } { \delta \eta }, \quad \eta _ { t } = \frac { \delta H } { \delta \zeta },\\
\zeta _ { 2, t } = - \frac { \delta H } { \delta \eta_2 }, \quad \eta _ { 2, t } = \frac { \delta H } { \delta \zeta_2 }.
\end{cases}
\end{equation}
Using the variables $q=\zeta _x$, $q_2=\zeta _{2,x}$, the system \eqref{6.39} can be represented in the form
\begin{equation}\label{6.40}
\begin{cases}
\eta _ { t } = - \left( \frac { \delta H } { \delta q } \right) _ { x },\quad  q _ { t } = - \left( \frac { \delta H } { \delta \eta } \right) _ { x },\\
\eta _ {2, t } = - \left( \frac { \delta H } { \delta q_2 } \right) _ { x },\quad
q _ { 2, t } = - \left( \frac { \delta H } { \delta \eta_2 } \right) _ { x }.
\end{cases}
\end{equation}
\section{Hamiltonian perturbation analysis}
\subsection{Approximation of the Dirichlet-Neumann Operators}
The expansion for the DN operators $G_1(\beta,\eta)$ is \cite{CGNS}
\begin{align}
G_1 ( \beta , \eta ) &=  D \tanh ( h_1 D ) + D L ( \beta ) + D \eta D - D \tanh ( h_1 D ) \eta D \tanh ( h_1 D ) \nonumber\\
\qquad\qquad\;\;\;&- D \tanh ( h_1 D ) \eta D L ( \beta ) - D L ( \beta ) \eta D \tanh ( h_1 D ) \nonumber\\
\qquad\qquad\;\;\;&- D L ( \beta ) \eta D L ( \beta ) + \mathcal { O } \left( \eta ^ { 2 } (h_1 D)^4\right),\label{6.42}
\end{align}
and the expansion of the operator $L ( \beta )$ is
\begin{equation*}
L ( \beta ) = \sum _ { j = 0 } ^ { \infty } L _ { j } ( \beta ),
\end{equation*}
where the operator $D=-i\partial/\partial x$, $L_ { j }\sim (\beta/h_1)^j$, and $|\beta(x)|/h_1<1$. Employing the recursive formulae in \cite{CGNS}, the first four terms of $L ( \beta )$ are calculated as
\begin{align}\label{6.43}
&L _ { 0 }  = 0, \nonumber\\
&L _ { 1 }  = - \operatorname { sech } ( h_1 D ) \beta D \operatorname { sech } ( h_1 D ), \nonumber\\
&L _ { 2 }  = - \operatorname { sech } ( h_1 D ) \beta D \sinh ( h_1 D ) \operatorname { sech } ( h_1 D ) \beta D \operatorname { sech } ( h_1 D ), \\
&L _ { 3 }= - \operatorname { sech } ( h_1 D )  \nonumber\\
&\qquad\times\left( \frac { \beta ^ { 3 } } { 3 ! } \operatorname { sech } ( h_1 D ) D ^ { 3 } + \frac { \beta ^ { 2 } } { 2 ! } D ^ { 2 } \cosh ( h_1 D ) L _ { 1 } - \frac { \beta } { 1 ! } D \sinh ( h_1 D ) L _ { 2 } \right).\nonumber
\end{align}
On the other hand, by virtue of the expansions
\begin{align}
&\tanh ( h_1 D ) = h_1 D - \frac { 1 } { 3 } h_1 ^ { 3 } D ^ { 3 } + \mathcal { O } \left( ( h_1 D ) ^ { 5 } \right),\nonumber\\
&\operatorname { sech } ( h_1 D ) = 1 - \frac { 1 } { 2 } h_1 ^ { 2 } D ^ { 2 } + \frac { 5 } { 24 } h_1 ^ { 4 } D ^ { 4 } + \mathcal { O } \left( h_1 ^ { 6 } D ^ { 6 } \right), \nonumber\\
&\sinh ( h_1 D ) = h_1 D + \frac { 1 } { 6 } h_1 ^ { 3 } D ^ { 3 } + \mathcal { O } \left( h_1 ^ { 5 } D ^ { 5 } \right), \nonumber\\
&\cosh ( h_1 D ) = 1 + \frac { 1 } { 2 } h_1 ^ { 2 } D ^ { 2 } + \frac { 1 } { 24 } h_1 ^ { 4 } D ^ { 4 } + \mathcal { O } \left( h_1 ^ { 6 } D ^ { 6 } \right),\nonumber
\end{align}
the expressions in \eqref{6.43} are formed as
\begin{equation*}
\begin{aligned}
L _ { 1 } & = - \left( 1 - \frac { 1 } { 2 } h_1 ^ { 2 } D ^ { 2 } + \mathcal { O } \left( h_1 ^ { 4 } D ^ { 4 } \right) \right) \beta D \left( 1 - \frac { 1 } { 2 } h_1 ^ { 2 } D ^ { 2 } + \mathcal { O } \left( h_1 ^ { 4 } D ^ { 4 } \right) \right) \\
& = - \beta D + \frac { 1 } { 2 } h_1 ^ { 2 } \beta D ^ { 3 } + \frac { 1 } { 2 } h_1 ^ { 2 } D ^ { 2 } \beta D + \mathcal { O } \left( h_1 ^ { 5 } D ^ { 5 } \right), \\
L _ { 2 } & = - \left( 1 + \mathcal { O } \left( h_1 ^ { 2 } D ^ { 2 } \right) \right) \beta D \left( h_1 D + \mathcal { O } \left( h_1 ^ { 3 } D ^ { 3 } \right) \right) \left( 1 + \mathcal { O } \left( h_1 ^ { 2 } D ^ { 2 } \right) \right) \beta D \left( 1 + \mathcal { O } \left( h_1 ^ { 2 } D ^ { 2 } \right) \right)\\
& = - h_1 \beta D ^ { 2 } \beta D + \mathcal { O } \left( h_1 ^ { 5 } D ^ { 5 } \right), \\
L _ { 3 } & = - \left( 1 + \mathcal { O } \left( h_1 ^ { 2 } D ^ { 2 } \right) \right) \left( \frac { \beta ^ { 3 } } { 3 ! } \Big( 1 + \mathcal { O } \left( h_1 ^ { 2 } D ^ { 2 } \right) \right) D ^ { 3 }\\
&\quad \;+ \frac { \beta ^ { 2 } } { 2 ! } D ^ { 2 } \left( 1 + \mathcal { O } \left( h_1 ^ { 2 } D ^ { 2 } \right) \right) \left( - \beta D + \mathcal { O } \left( h_1 ^ { 3 } D ^ { 3 } \right) \right) \\
&\quad \;- \beta D \left( h_1 D + \mathcal { O } \left( h_1 ^ { 3 } D ^ { 3 } \right) \right) \left( - \beta D^2 \beta ( h_1 D ) + \mathcal { O } \left( h_1 ^ { 5 } D ^ { 5 } \right) \right)\Big) \\
& = - \frac { 1 } { 6 } \beta ^ { 3 } D ^ { 3 } + \frac { 1 } { 2 } \beta ^ { 2 } D ^ { 2 } \beta D + \mathcal { O } \left( h_1 ^ { 5 } D ^ { 5 } \right).
\end{aligned}
\end{equation*}
Thus the truncated expansion of $G _1( \beta , \eta )$  in \eqref{6.42} is
\begin{equation}\label{4.3-1}
\begin{aligned}
G _1( \beta , \eta ) &= D \Big( h_1 - \beta - \frac { h _1^ { 3 } } { 3 } D ^ { 2 } + \frac { h_1 ^ { 2 } } { 2 } \beta D ^ { 2 } + \frac { h_1 ^ { 2 } } { 2 } D ^ { 2 } \beta - h_1 \beta D ^ { 2 } \beta \\
&\quad\;- \frac { 1 } { 6 } \beta ^ { 3 } D ^ { 2 } + \frac { 1 } { 2 } \beta ^ { 2 } D ^ { 2 } \beta + \eta \Big) D + \mathcal { O } \left( \eta (h_1 D)^4 , ( h_1 D ) ^ { 6 } \right).
\end{aligned}
\end{equation}

In terms of Fourier multipliers the leading order terms of the Dirichlet-Neumann operator $G _2( \eta,\eta_2 )$ are given by \cite{CGK}
\begin{align}\label{4.4-1}
G _{11}(\eta,\eta_2 )=& D \coth\left(h_{2} D\right)+D \coth\left(h_{2} D\right) \eta(x) D \coth\left(h_{2} D\right)-D \eta(x) D \nonumber\\
&-D \operatorname{csch}\left(h_{2} D\right) \eta_{2}(x) D \operatorname{csch}\left(h_{2} D\right),\nonumber\\
G _{12}(\eta,\eta_2 )=&-D \operatorname{csch}\left(h_{2} D\right)-D\operatorname{coth}\left(h_{2} D\right) \eta(x) D \operatorname{csch}\left(h_{2} D\right)\nonumber\\
&+D\operatorname{csch}\left(h_{2} D\right) \eta_{2}(x) D \operatorname{coth}\left(h_{2} D\right),\nonumber\\
G _{22}(\eta,\eta_2 )=& D \operatorname{coth}\left(h_{2} D\right) + D \operatorname{csch}\left(h_{2} D\right) \eta(x) D \operatorname{csch}\left(h_{2} D\right)\nonumber\\
&-D \operatorname{coth}\left(h_{2} D\right) \eta_{2}(x) D \operatorname{coth}\left(h_{2} D\right)+D \eta_{2}(x) D,
\end{align}
and $G_{21}^{*}=G_{12}$. By virtue of the expansions
\begin{align}
&\operatorname { csch } ( h_2 D ) = ( h_2 D )^{-1}-\frac{1}{6} ( h_2 D )+\frac{7}{360} ( h_2 D )^{3}+\mathcal { O } \left(( h_2 D )^{5}\right), \nonumber\\
&\coth ( h_2 D ) = ( h_2 D )^{-1}+\frac{1}{3}  ( h_2 D )-\frac{1}{45}  ( h_2 D )^{3}+\mathcal { O } \left( ( h_2 D )^{5}\right),\nonumber
\end{align}
the truncated expansions in \eqref{4.4-1} are formed as
\begin{align}
G _{11}(\eta,\eta_2 )=&\frac{1}{h_{2}}+\left\{\frac{1}{3} h_{2}D^{2}+\frac{1}{h_{2}^{2}} \eta(x)\right\}-\frac{1}{h_{2}^{2}} \eta_{2}(x)\nonumber\\
&+\left\{-\frac{1}{45} h_{2}^{3}(D)^{4}+\frac{1}{3}\eta(x)(D)^{2}
+\frac{1}{3}(D)^{2} \eta(x)-D \eta D\right\}\nonumber\\
&+\left\{\frac{1}{6} \eta_{2}(x)(D)^{2}+\frac{1}{6}(D)^{2} \eta_{2}\left(x\right)\right\}+\mathcal { O } \left(\eta h_2^2D^4,\eta_2 h_2^2D^4\right),\label{4.5-1}\\
G _{12}(\eta,\eta_2 )=&-\frac{1}{h_{2}}+\left\{-\frac{1}{h_{2}^{2}} \eta+\frac{h_{2}}{6}\left(D\right)^{2}\right\}+\mathcal { O } \left(\frac 1 {h_{2}^2}\eta_2,h_2^3D^4\right),\label{4.6-1}\\
G _{21}(\eta,\eta_2 )=&-\frac{1}{h_{2}}+\left\{-\frac{1}{h_{2}^{2}} \eta+\frac{h_{2}}{6}\left(D\right)^{2}\right\}+\mathcal { O } \left(\frac 1 {h_{2}^2}\eta_2,h_2^3D^4\right),\label{4.7-1}\\
G _{22}(\eta,\eta_2 )=&\frac{1}{h_{2}}+\left\{\frac{1}{h_{2}^{2}} \eta+\frac{h_{2}}{3}\left(D\right)^{2}\right\}+\mathcal { O } \left(\frac 1 {h_{2}^2}\eta_2,h_2^3D^4\right).\label{4.8-1}
\end{align}
\subsection{Scales and approximations}
We introduce the small parameters $\varepsilon\ll1$ and $\delta\ll1$ and consider the shallow-water (long wave) regime defined by the spatial scale \cite{CoI19}
\begin{equation}\label{2.20}
x^{\prime}=\varepsilon x,
\end{equation}
and the wave perturbation scales
\begin{equation}\label{2.21}
\begin{cases}
\eta(x, t)=\varepsilon^{2} \eta^{\prime}\left(x^{\prime}, t\right), \quad q(x, t)=\varepsilon^{2} q^{\prime}\left(x^{\prime}, t\right), \\
\eta_{2}(x, t)=\delta \varepsilon^{2} \eta_{2}^{\prime}\left(x^{\prime}, \frac{t}{\delta}\right),\; q_{2}(x, t)=\delta \varepsilon^{2} q_{2}^{\prime}\left(x^{\prime}, \frac{t}{\delta}\right),\\
\beta(x)=\varepsilon^2\beta'(x'),
\end{cases}
\end{equation}
where $ \eta $ and $ q $, as well as $ \eta_{2} $ and $ q_{2} $ are considered of the same order of magnitude, with the surface wave having a smaller amplitude than the internal wave and the quantities $ x^{\prime}, t, \eta^{\prime}, q^{\prime}, \eta_{2}^{\prime} $, $ q_{2}^{\prime} $ and $ \beta' $ are all of order $ \mathcal { O }(1) $. The interpretation of the specific scaling \eqref{2.21} can be referred to \cite{CoI19}.
The scaling \eqref{2.20}-\eqref{2.21} transforms the system \eqref{6.40} into the Hamiltonian system
\begin{equation}\label{4.16-1}
\begin{cases}
\eta' _ { t } = - \left( \frac { \delta H' } { \delta q' } \right) _ { x' },\quad  q' _ { t } = - \left( \frac { \delta H' } { \delta \eta' } \right) _ { x' },\\
\eta' _ {2, t } = - \left( \frac { \delta H' } { \delta q'_2 } \right) _ { x' },\quad
q' _ { 2, t } = - \left( \frac { \delta H' } { \delta \eta'_2 } \right) _ { x' },
\end{cases}
\end{equation}
with the Hamiltonian $H'=H\varepsilon^{-3}$, and $\xi=\varepsilon\xi', \xi_2=\varepsilon\delta\xi'_2$ for
\begin{equation}\label{4.17-1}
\partial_{x^{\prime}} \xi^{\prime}=q^{\prime}-\frac \Gamma 2 \eta^{\prime}, \quad \partial_{x^{\prime}} \xi_{2}^{\prime}=q_{2}^{\prime}-\frac {\Gamma_{2}} 2 \eta_{2}^{\prime}.
\end{equation}
Denoting $b^{\prime}\left(x^{\prime}\right)= h_1 - \varepsilon^{2}\beta'(x')$,
we can interchange $ b' D' $ and $ D' b' $ as the difference $ D' b'- b' D' \sim \varepsilon^2$ and could be neglected. Then the DN operators in \eqref{4.3-1} and \eqref{4.5-1}-\eqref{4.8-1} can be represented in the form
\begin{align}
G _1( b^{\prime} , \eta^{\prime} )=& \varepsilon^2 D' \left( b^{\prime}\left(x^{\prime}\right) - \frac { 1 } { 3 } \varepsilon^2 D' (b^{\prime}\left(x^{\prime}\right))^ { 3 } D' + \varepsilon^2 \eta' \left(x^{\prime}\right)\right) D'+ \mathcal { O } \left( \varepsilon ^ { 6 } \right),\label{2.23}\\
G _{11}(\eta',\eta_2' )=&\frac{1}{h_{2}}+\varepsilon^2\left\{\frac{1}{3} h_{2}(D')^{2}+\frac{1}{h_{2}^{2}} \eta'(x')\right\}-\varepsilon^2\delta\frac{1}{h_{2}^{2}} \eta'_{2}(x')\nonumber\\
&+\varepsilon^4\left\{-\frac{1}{45} h_{2}^{3}(D')^{4}+\frac{1}{3}\eta'(x')(D')^{2}
+\frac{1}{3}(D')^{2} \eta'(x')-D' \eta'(x') D'\right\}\nonumber\\
&+\varepsilon^4\delta\left\{\frac{1}{6} \eta'_{2}(x')(D')^{2}+\frac{1}{6}(D')^{2} \eta'_{2}(x')\right\}+\mathcal { O } \left(\varepsilon^6\right),\label{4.12-1}\\
G _{12}(\eta',\eta'_2 )=&-\frac{1}{h_{2}}+\varepsilon^2\left\{-\frac{1}{h_{2}^{2}} \eta'+\frac{h_{2}}{6}(D')^{2}\right\}+\mathcal { O } \left(\varepsilon^2\delta,\varepsilon^4\right),\label{4.13-1}\\
G _{21}(\eta',\eta'_2 )=&-\frac{1}{h_{2}}+\varepsilon^2\left\{-\frac{1}{h_{2}^{2}} \eta'+\frac{h_{2}}{6}(D')^{2}\right\}+\mathcal { O } \left(\varepsilon^2\delta,\varepsilon^4\right),\label{4.14-1}\\
G _{22}(\eta',\eta'_2 )=&\frac{1}{h_{2}}+\varepsilon^2\left\{\frac{1}{h_{2}^{2}} \eta'+\frac{h_{2}}{3}(D')^{2}\right\}+\mathcal { O } \left(\varepsilon^2\delta,\varepsilon^4\right).\label{4.15-1}
\end{align}
The operator $\mathcal{B}=\rho_{1} G_{11}+\rho_{2} G_{1}$, which is a function of DN operators, can thus be transformed as
\begin{align}
\mathcal{B}(b',\eta',\eta'_2) &=\frac{\rho_{1}}{h_{2}}+\varepsilon^2\left\{\rho_{2}D' b'(x')D'+\frac{\rho_{1}}{3}h_{2}(D')^{2}+\frac{\rho_{1}}{h_{2}^{2}} \eta'(x')\right\}\nonumber\\
&-\varepsilon^2\delta\frac{\rho_{1}}{h_{2}^{2}} \eta'_{2}(x')-\varepsilon^4\Big\{\frac { \rho_{2} } { 3 }(D')^2 (b'(x'))^ { 3 } (D')^2 - \rho_{2} D'\eta'(x') D'
+\frac{\rho_{1}}{45} h_{2}^{3}(D')^{4}\nonumber\\
&-\frac{\rho_{1}}{3}\eta'(x')(D')^{2}-\frac{\rho_{1}}{3}(D')^{2} \eta'(x')+\rho_{1}D'\eta'(x') D'\Big\}+\mathcal { O } \left(\varepsilon^4\delta,\varepsilon^6\right).
\end{align}
Then
\begin{align}\label{2.25}
[\mathcal{B}(b',\eta',\eta'_2)]^{-1}
=&\frac{h_{2}}{\rho_{1}}-\frac{\varepsilon^2h^2_{2}}{\rho^2_{1}}\left\{\rho_{2}D' b'(x')D'+\frac{\rho_{1}}{3}h_{2}(D')^{2}+\frac{\rho_{1}}{h_{2}^{2}} \eta'(x')\right\}\nonumber\\
&+\frac{\varepsilon^2\delta}{\rho_{1}} \eta'_{2}(x')+\mathcal { O } \left(\varepsilon^4\right),
\end{align}
where the expansion $(a+x)^{-1}=\frac 1 a - \frac 1 {a^2}x + \frac 1 {a^3}x^{2}+\mathcal{O}\left(x^{3}\right)$ is used.

Choosing $ \delta=O(\varepsilon) $ in \eqref{2.21}, we now write the Hamiltonian \eqref{6.24} as functionally dependent on $ \eta^{\prime}, \eta_{2}^{\prime}, \xi^{\prime}, \xi_{2}^{\prime}, b' $ by retaining terms up to order $ O\left(\varepsilon^{5}\right) $.
By \eqref{2.23}-\eqref{2.25}, the first term of \eqref{6.24} can be written as
\begin{align}\label{2.26}
\frac{ \varepsilon^2\delta^2}{2\rho_2} \int_{\mathbb{R}} \xi'_{2}\left\{ \frac{1}{h_{2}}+\varepsilon^2\left(\frac{1}{h_{2}^{2}} \eta'+\frac{h_{2}}{3}(D')^{2}\right)\right\} \xi'_{2} \frac {d x'}{\varepsilon}
+\mathcal { O } \left(\varepsilon^6\right).
\end{align}
As
\begin{align*}
G_1 \mathcal{B}^{-1} G_{11}=& \varepsilon^{2}\frac{1}{\rho_1}\left(D'b'D'\right)
+\varepsilon^{4}\Big\{
-\frac{1}{3\rho_1}\left(D^{\prime}\right)^{2}(b')^3\left(D^{\prime}\right)^{2}\\
&+\frac{1}{\rho_1} D^{\prime} \eta^{\prime}\left(x^{\prime}\right) D^{\prime}-\frac{\rho_{2} h_{2}}{(\rho_1)^{2}}\left(D^{\prime}\right)^{2}(b')^2\left(D^{\prime}\right)^{2}\Big\}
+O\left(\varepsilon^{5}\right),
\end{align*}
the second term of \eqref{6.24} can be written as
\begin{align}\label{2.27}
&\frac{\varepsilon^4}{2} \int_{\mathbb{R}}\xi' \Big\{\frac{1}{\rho_1}\left(D'b'D'\right)+
\varepsilon^{2}\big(
-\frac{1}{3\rho_1}\left(D^{\prime}\right)^{2}(b')^3\left(D^{\prime}\right)^{2}\nonumber\\
&+\frac{1}{\rho_1} D^{\prime} \eta^{\prime}\left(x^{\prime}\right) D^{\prime}-\frac{\rho_{2} h_{2}}{(\rho_1)^{2}}\left(D^{\prime}\right)^{2}(b')^2\left(D^{\prime}\right)^{2}\big)\Big\} \xi'\frac {d x'}{\varepsilon}+\mathcal { O } \left(\varepsilon^6 \right).
\end{align}
The sum of third and fourth terms of \eqref{6.24} can be written as
\begin{equation}\label{2.28}
\frac{\varepsilon^4\delta}{\rho_1} \int_{\mathbb{R}}\xi'D'b'D'\xi_2'\frac {d x'}{\varepsilon}+\mathcal { O } \left(\varepsilon^6 \right).
\end{equation}
As
\begin{align*}
G_{21} \mathcal{B}^{-1} G_{12}
=\frac{1}{\rho_1 h_{2}}+\varepsilon^{2}\left\{\frac{1}{\rho_1 h_{2}^{2}} \eta^{\prime}-\frac{2 h_{2}}{3\rho_1}\left(D^{\prime}\right)^{2}-\frac{\rho_2}{\rho_1^{2}}
D'b'D'\right\}+O\left(\varepsilon^{3}\right),
\end{align*}
the fifth term of \eqref{6.24} can be written as
\begin{align}\label{2.29}
-\frac{\varepsilon^{2}\delta^2}{2\rho_2} \int_{\mathbb{R}} \xi'_{2}
\left(\frac{1}{h_{2}}+\varepsilon^{2}\left\{\frac{1}{ h_{2}^{2}} \eta^{\prime}-\frac{2 h_{2}}{3}\left(D^{\prime}\right)^{2}-\frac{\rho_2}{\rho_1}
D'b'D'\right\}\right)  \xi'_{2}
\frac {d x'}{\varepsilon}+\mathcal { O } \left(\varepsilon^6 \right).
\end{align}
Adding \eqref{2.26} and \eqref{2.29}, we obtain
\begin{equation}\label{2.30}
\frac{\varepsilon^{4}\delta^2}{2} \left(\frac{h_{2}}{\rho_2}\int_{\mathbb{R}}
\xi'_{2}(D')^2\xi'_{2}\frac {d x'}{\varepsilon}+\frac{1}{\rho_1}
\int_{\mathbb{R}}\xi'_{2}D'b'D'\xi'_{2}\frac {d x'}{\varepsilon}\right)
+\mathcal { O } \left(\varepsilon^6 \right).
\end{equation}
Using \eqref{2.27}, \eqref{2.28}, \eqref{2.30}, and retaining only $\mathcal { O } \left( \varepsilon^5\right)$ terms, the Hamiltonian \eqref{6.24} is approximated as
\begin{align}\label{4.25-1}
&H^{(5)}=\frac{\varepsilon^{4}\delta^2}{2} \left(\frac{h_{2}}{\rho_2}\int_{\mathbb{R}}
\xi'_{2}(D')^2\xi'_{2}\frac {d x'}{\varepsilon}+\frac{1}{\rho_1}
\int_{\mathbb{R}}\xi'_{2}D'b'D'\xi'_{2}\frac {d x'}{\varepsilon}\right)
+\frac{\varepsilon^4}{2\rho_1} \int_{\mathbb{R}}\xi'\left(D'b'D'\right)\xi'\frac {d x'}{\varepsilon}\nonumber\\
&+\frac{\varepsilon^6}{2\rho_1} \int_{\mathbb{R}} \xi'\Big\{
-\frac{1}{3}\left(D^{\prime}\right)^{2}(b')^3\left(D^{\prime}\right)^{2}
+D^{\prime} \eta^{\prime}\left(x^{\prime}\right) D^{\prime}-\frac{\rho_{2} h_{2}}{\rho_1}\left(D^{\prime}\right)^{2}(b')^2\left(D^{\prime}\right)^{2}\Big\} \xi'\frac {d x'}{\varepsilon}\nonumber\\
&+\frac{\varepsilon^4\delta}{\rho_1} \int_{\mathbb{R}}\xi'D'b'D'\xi_2'\frac {d x'}{\varepsilon}+\varepsilon^4 \frac { g \left( \rho _ { 1 } - \rho _ { 2 } \right)+\kappa\left( \rho _ { 1 } \gamma _ { 1}- \rho _ { 2 } \gamma _ { 2}\right) } { 2 }  \int _ { \mathbb { R } } (\eta') ^ { 2 } \frac {d x'}{\varepsilon}\nonumber\\
&+\varepsilon^4 \delta^2\frac { g\rho _ { 2 }+\rho _ { 2 } \gamma _ { 2 } ( \gamma _ { 2 } h_2+\kappa) } { 2} \int _ { \mathbb { R } } (\eta' _ { 2 }) ^ { 2 } \frac {d x'}{\varepsilon}
+\frac {\varepsilon^6 ( \rho _ { 1 } \gamma _ { 1}^2-\rho _ { 2 } \gamma _ { 2}^2) } { 6 } \int _ { \mathbb { R } } \eta ^ { 3 } \frac {d x'}{\varepsilon}\nonumber\\
&-\varepsilon^4 \delta^2 \left( \gamma _ { 2 }h_2 +\kappa\right)\int _ { \mathbb { R } } \xi'_2\eta' _ {2, x' }\frac {d x'}{\varepsilon}
-\varepsilon^4\int_{\mathbb{R}} (\varepsilon^2\gamma_{1} \eta'+\kappa) \eta'_{x'}\xi'
\frac {d x'}{\varepsilon}.
\end{align}
On the other hand, the relations \eqref{4.17-1} and the definition of $b'(x')$ yield
\begin{align}\label{2.31}
&H^{(5)}=\frac{\varepsilon^{3}}{2} \int_{\mathbb{R}}\left(\frac{b'}{\rho_1} q^{\prime 2}+(2\kappa- \frac{\Gamma b'}{\rho_1}) q^{\prime} \eta^{\prime}+A_{1} \eta^{\prime 2}+\delta^{2}\rho_2\left(g-2\omega \left( \gamma _ { 2 }h_2 +\kappa\right)\right)\left(\eta_{2}^{\prime}\right)^{2}\right) d x^{\prime}\nonumber\\
&+\varepsilon^{3} \delta \int_{\mathbb{R}}\frac{b'}{\rho_1}\left(q^{\prime} q_{2}^{\prime}-\frac {\Gamma} 2 \eta^{\prime} q_{2}^{\prime}-\frac {\Gamma_{2}} 2 \eta_{2}^{\prime} q^{\prime}+\frac {\Gamma \Gamma_2} 4 \eta^{\prime} \eta_{2}^{\prime}\right) d x^{\prime}
+\varepsilon^{3} \delta^{2} \left( \gamma _ { 2 }h_2 +\kappa\right) \int_{\mathbb{R}} q_{2}^{\prime} \eta_{2}^{\prime} d x^{\prime} \nonumber\\
&-\varepsilon^{5} \int_{\mathbb{R}} \frac{A_{2}}{2} \left(\left(q_{x}^{\prime}\right)^{2}-\Gamma q_{x^{\prime}}^{\prime} \eta_{x^{\prime}}^{\prime}+\frac {\Gamma^{2}} 4\left(\eta_{x^{\prime}}^{\prime}\right)^{2}\right) d x^{\prime}\nonumber\\
&+\varepsilon^{5} \frac{1}{2\rho_1} \int_{\mathbb{R}} \eta^{\prime}\left(q^{\prime}\right)^{2} d x^{\prime}
+\varepsilon^{5}\left(\frac{\gamma_1}{2}-\frac{\Gamma}{2\rho_1}\right) \int_{\mathbb{R}} q^{\prime}\left(\eta^{\prime}\right)^{2} d x^{\prime}
+\varepsilon^{5} \frac{A_{3}}{6} \int_{\mathbb{R}}\left(\eta^{\prime}\right)^{3} d x^{\prime}\nonumber\\
&+\frac{\varepsilon^3 \delta^{2}}{2}
\int_{\mr}\left(\frac{b'}{\rho_1}+\frac{h_2}{\rho_2}\right)
\left(\left(q_{2}^{\prime}\right)^{2}- \Gamma_{2} q_{2}^{\prime} \eta_{2}^{\prime}+\frac {\Gamma_{2}^{2}} 4\left(\eta_{2}^{\prime}\right)^{2}\right) d x^{\prime},
\end{align}
where
\begin{align}\label{2.32}
&A_{1}(x')= \left( \rho _ { 1 } - \rho _ { 2 } \right)\left( g-2\omega \kappa\right) +\frac{b' \Gamma^{2}}{4\rho_1},\nonumber\\
&A_{2}=\frac{(h_1)^{2}\left[\rho_1h_1+3 \rho_2 h_{2}\right]}{3(\rho_1)^{2}},\nonumber\\
&A_{3}=\rho_1 \gamma_1^{2}-\rho_2\gamma_{2}^{2}-\frac {3 \gamma_1\Gamma} 2+\frac{3 \Gamma^{2}}{4\rho_1}.
\end{align}
\section{KdV-type equation describing the evolution of the internal waves}
We truncate the Hamiltonian \eqref{6.24} at $\mathcal { O } \left( \varepsilon^6\right) $ and use \eqref{4.16-1} to reach that the canonical equations for the variables related to the free surface are linear
\begin{align}
\eta_{2, t}^{\prime}=&-\partial_{x^{\prime}}\left[\delta \left( \gamma _ { 2 }h_2 +\kappa\right) \eta_{2}^{\prime}+\frac{b'}{\rho_1}\left(q^{\prime}-\frac {\Gamma} 2 \eta^{\prime}\right)+\delta\left(\frac{b'}{\rho_1}+\frac{h_2}{\rho_2}\right)
\left(q_{2}^{\prime}-\frac {\Gamma_2} 2 ^{\prime} \eta_{2}^{\prime}\right)\right]\label{3.1}\\
q_{2, t}^{\prime}=&-\partial_{x^{\prime}}\left[\delta \left( \gamma _ { 2 }h_2 +\kappa\right) q_{2}^{\prime}-\delta \frac {\Gamma_2} 2\left(\frac{b'}{\rho_1}+\frac{h_2}{\rho_2}\right)
\left(q_{2}^{\prime}-\frac {\Gamma_2} 2 \eta_{2}^{\prime}\right)\right.\nonumber\\
&\left.+\delta\rho_2\left(g-2\omega \left( \gamma _ { 2 }h_2 +\kappa\right)\right) \eta_{2}^{\prime}-\frac {\Gamma_2} 2\frac{b'}{\rho_1}\left(q^{\prime}-\frac {\Gamma} 2 \eta^{\prime}\right)\right],\label{3.2}
\end{align}
and those for the variables at the internal waves are nonlinear
\begin{align}
\eta_{t}^{\prime}=&-\partial_{x^{\prime}}\left[\kappa\eta^{\prime}
+\frac{b'}{\rho_1}\left(q^{\prime}-\frac {\Gamma} 2
\eta^{\prime}+\delta\left(q_{2}^{\prime}-\frac {\Gamma_2} 2 \eta_{2}^{\prime}\right)\right)+\varepsilon^{2} A_{2}\left(q_{x^{\prime} x^{\prime}}^{\prime}-\frac {\Gamma} 2 \eta_{x^{\prime} x^{\prime}}^{\prime}\right)\right.\nonumber\\ &\left.+\varepsilon^{2} \frac{1}{\rho_1} q^{\prime} \eta^{\prime}
+\varepsilon^{2}\left(\frac{\gamma_1}{2}-\frac{\Gamma}{2\rho_1}\right) \eta^{\prime 2}\right],\label{3.3}\\
q_{t}^{\prime}=&-\partial_{x^{\prime}}\left[\kappa q^{\prime}-\frac{\Gamma b'}{2\rho_1} q^{\prime}+A_{1}(x') \eta^{\prime}-\delta \frac{\Gamma b'}{2\rho_1}\left(q_{2}^{\prime}-\frac {\Gamma_2} 2 \eta_{2}^{\prime}\right)-\varepsilon^{2} A_{2} \frac{\Gamma}{2}\left(q_{x^{\prime} x^{\prime}}^{\prime}-\frac{\Gamma}{2} \eta_{x^{\prime} x^{\prime}}^{\prime}\right)\right.\nonumber\\ &\left.+\varepsilon^{2} \frac{1}{\rho_1} \frac{q^{\prime 2}}{2}+\varepsilon^{2}\left(\gamma_1-\frac{\Gamma}{\rho_1}\right)\eta^{\prime} q^{\prime}+\varepsilon^{2} A_{3} \frac{\eta^{\prime 2}}{2}\right].\label{3.4}
\end{align}
Considering $ \delta \ll \varepsilon^{2} $, we neglect the $ \delta $-terms above and the
system \eqref{3.1}-\eqref{3.4} is reduced to
\begin{align}
\eta_{2, t}^{\prime}=&-\partial_{x^{\prime}}\left[\frac{b'}{\rho_1}\left(q^{\prime}-\frac {\Gamma} 2 \eta^{\prime}\right)\right],\label{3.5}\\
q_{2, t}^{\prime}=&\frac {\Gamma_2} 2\partial_{x^{\prime}}\left[\frac{b'}{\rho_1}\left(q^{\prime}-\frac {\Gamma} 2 \eta^{\prime}\right)\right],\label{3.6}\\
\eta_{t}^{\prime}=&-\partial_{x^{\prime}}\left[\kappa\eta^{\prime}
+\frac{b'}{\rho_1}\left(q^{\prime}-\frac {\Gamma} 2
\eta^{\prime}\right)+\varepsilon^{2} A_{2}\left(q_{x^{\prime} x^{\prime}}^{\prime}-\frac {\Gamma} 2 \eta_{x^{\prime} x^{\prime}}^{\prime}\right)\right.\nonumber\\ &\left.+\varepsilon^{2} \frac{1}{\rho_1} q^{\prime} \eta^{\prime}
+\varepsilon^{2}\left(\frac{\gamma_1}{2}-\frac{\Gamma}{2\rho_1}\right) \eta^{\prime 2}\right],\label{3.7}\\
q_{t}^{\prime}=&-\partial_{x^{\prime}}\left[(\kappa-\frac{\Gamma b'}{2\rho_1}) q^{\prime}+A_{1}(x') \eta^{\prime}-\varepsilon^{2} A_{2} \frac{\Gamma}{2}\left(q_{x^{\prime} x^{\prime}}^{\prime}-\frac{\Gamma}{2} \eta_{x^{\prime} x^{\prime}}^{\prime}\right)\right.\nonumber\\ &\left.+\varepsilon^{2} \frac{1}{\rho_1} \frac{q^{\prime 2}}{2}+\varepsilon^{2}\left(\gamma_1-\frac{\Gamma}{\rho_1}\right)\eta^{\prime} q^{\prime}+\varepsilon^{2} A_{3} \frac{\eta^{\prime 2}}{2}\right],\label{3.8}
\end{align}
which shows that the internal wave motion is decoupled from the oscillations of the free surface. From \eqref{3.5} and \eqref{3.6}, we note that the motion of the free surface is determined by the initial data as $\left(q_{2}^{\prime}+\frac {\Gamma_2} 2 \eta_{2}^{\prime}\right)_{t}=0 $ and by the characteristics $(\eta',q')$ of the displacements of the internal waves and by the characteristics $ b'$ of the bottom variation. On the other hand, the equations \eqref{3.7} and \eqref{3.8} yield that the leading order linear equations for $(\eta',q')$ satisfy
\begin{align}
\eta_{t}^{\prime}=&-\partial_{x^{\prime}}\left[(\kappa-\frac{\Gamma b'}{2\rho_1}) \eta^{\prime}+\frac{b'}{\rho_1}q^{\prime}\right],\label{3.9}\\
q_{t}^{\prime}=&-\partial_{x^{\prime}}\left[(\kappa-\frac{\Gamma b'}{2\rho_1}) q^{\prime}+A_{1} \eta^{\prime}\right].\label{3.10}
\end{align}
To derive the solutions for $\eta'$ and $q'$ of \eqref{3.9} and \eqref{3.10}, we take the form
\begin{align}\label{3.11}
&\eta' ( x' , t ) = \eta' _ { 0 } e ^ { i k ( x' - c ( x' ) t ) }, \nonumber\\
&q' ( x' , t ) = q' _ { 0 } e ^ { i k ( x' - c ( x' ) t ) },
\end{align}
with $k$ being the wave number and $c(x')$ being the wave speed dependent on variable $x'$.
Substituting \eqref{3.11} into \eqref{3.9} and \eqref{3.10} yields
\begin{align}
&-c\eta'+ (\kappa-\frac{\Gamma b'}{2\rho_1}) \eta^{\prime}+\frac{b'}{\rho_1}q^{\prime}  =0, \label{3.12}\\
&-cq' + (\kappa-\frac{\Gamma b'}{2\rho_1}) q^{\prime}+A_{1} \eta^{\prime} =0.\label{3.13}
\end{align}
From \eqref{3.12}
\begin{equation}\label{3.14}
q'=\frac{\rho_1}{b'}\left(c-\kappa+\frac{\Gamma b'}{2\rho_1}\right) \eta^{\prime},
\end{equation}
and so inserting this into \eqref{3.13} gives the following quadratic expression for the wave speed $c$
\begin{equation}\label{3.15}
\left(c-\kappa+\frac{\Gamma b'}{2\rho_1}\right)^2-\frac{b'\left( \rho _ { 1 } - \rho _ { 2 } \right)\left( g-2\omega \kappa\right)}{\rho_1} -\frac{(b')^2 \Gamma^{2}}{4(\rho_1)^2} = 0.
\end{equation}
The solutions are
\begin{equation}\label{3.16}
c(x')=\kappa-\frac{\Gamma b'}{2\rho_1} \pm \sqrt{\frac{(\Gamma b')^2 }{(2\rho_1)^2}+\frac{b'\left( \rho _ { 1 } - \rho _ { 2 } \right)\left( g-2\omega \kappa\right)}{\rho_1} }.
\end{equation}
The solutions of \eqref{3.16} are real as $ g \gg 2\omega \kappa $ and $\rho_1>\rho_2$.
Besides, recalling from \eqref{2.5} that $ \kappa $ is the speed of the underlying current at the mean level $ z=0 $ of the internal waves, we obtain that in \eqref{3.16} that the plus sign corresponds to the speed of waves outrunning the current (downstream linear waves), while the minus sign corresponds to the speed of waves running counter to the current (upstream linear waves).
\begin{remark}
Setting $b'=h_1$, i.e. $\beta=0$, the  dispersion relation \eqref{3.16} recovers the result in \cite{CoI19} for fluids with flat bottom and for the special case $h_2=0$ (i.e., taking $\gamma_2=\rho_2=0$), the equation \eqref{3.16} recovers the dispersion relation in the single-layer case \cite{CIMT}.
\end{remark}
Inspired by the previous studies \cite{CIMT,CIT,IMT}, our next step is to transfer the system \eqref{3.7}-\eqref{3.8} to the KdV-type equation. Setting
\begin{equation}\label{3.18}
\begin{cases}
c_{0}(x')=c(x')-\kappa,\\
c_{1}(x')=c(x')-\kappa+\frac{\Gamma b'}{2\rho_1}
=\pm \sqrt{\frac{(\Gamma b')^2 }{(2\rho_1)^2}+\frac{b'\left( \rho _ { 1 } - \rho _ { 2 } \right)\left( g-2\omega \kappa\right)}{\rho_1} }=\pm \sqrt{\frac {b'} {\rho_1}A_1},
\end{cases}
\end{equation}
then \eqref{3.14} gives that
\begin{equation}\label{3.17}
q'=\frac{\rho_1}{b'}c_{1}(x')\eta^{\prime}.
\end{equation}
We introduce the additional characteristic variable
\begin{equation*}
\theta = \frac { 1 } { \varepsilon^2 } R ( X ) - t,
\end{equation*}
where $ R ( X )$ is a function such as $R ^ { \prime } ( X ) \equiv \frac { 1 } { c ( X ) }$ and $X=\varepsilon^2x'$. Our next goal is to transform the system \eqref{3.7}-\eqref{3.8} from variables $(x',t)$ to the slow variables $(X,\theta)$. To this end, we calculate straightforward the relations between the derivatives as
\begin{equation*}
\partial _ { x' } \equiv \frac { 1 } { c ( X ) } \partial _ { \theta } + \varepsilon^2 \partial _ { X },\qquad
\partial _ { t } \equiv - \partial _ { \theta }.
\end{equation*}
The system of \eqref{3.7} and \eqref{3.8} are therefore rewritten in terms of $(\theta,X)$ as
\begin{align}\label{3.19}
&-c\eta_{\theta}^{\prime}+\partial_{\theta}\left((\kappa-\frac{\Gamma b'}{2\rho_1})
\eta^{\prime}+\frac{b'}{\rho_1}q^{\prime}\right)
+\varepsilon^{2}c\partial_{X}\left((\kappa-\frac{\Gamma b'}{2\rho_1})\eta^{\prime}
+\frac{b'}{\rho_1}q^{\prime}\right)\nonumber\\
&=-\varepsilon^{2}\partial_{\theta}\left[ \frac {A_{2}}{c^2}\left(q_{\theta \theta}^{\prime}-\frac {\Gamma} 2 \eta_{\theta \theta}^{\prime}\right)+\frac{1}{\rho_1} q^{\prime} \eta^{\prime}
+\left(\frac{\gamma_1}{2}-\frac{\Gamma}{2\rho_1}\right) \eta^{\prime 2}\right]
+ \mathcal { O } \left( \varepsilon ^ { 4 } \right),
\end{align}
and
\begin{align}\label{3.20}
&-cq_{\theta}^{\prime}+\partial_{\theta}\left((\kappa-\frac{\Gamma b'}{2\rho_1}) q^{\prime}+A_{1} \eta^{\prime}\right)
+\varepsilon^{2}c\partial_{X}\left((\kappa-\frac{\Gamma b'}{2\rho_1}) q^{\prime}+A_{1} \eta^{\prime}\right)\nonumber\\
&=\varepsilon^{2}\partial_{\theta}\left[ \frac {A_{2}}{c^2} \frac{\Gamma}{2}\left(q_{\theta \theta}^{\prime}-\frac{\Gamma}{2} \eta_{\theta \theta}^{\prime}\right)- \frac{1}{\rho_1} \frac{q^{\prime2}}{2}
-\left(\gamma_1-\frac{\Gamma}{\rho_1}\right)\eta^{\prime} q^{\prime}- A_{3} \frac{\eta^{\prime 2}}{2}\right]+\mathcal { O } \left( \varepsilon ^ { 4 } \right).
\end{align}
From \eqref{3.19},
\begin{align}
q' _ { \theta } = &\frac{\rho_1} {b'} ( c - \kappa+ \frac{\Gamma b'}{2\rho_1}) \eta' _ { \theta } - \varepsilon^{2}\frac{\rho_1c} {b'}\partial_{X}\left((\kappa-\frac{\Gamma b'}{2\rho_1})\eta^{\prime}
+\frac{b'}{\rho_1}q^{\prime}\right)\nonumber\\
&-\varepsilon^{2}\frac{\rho_1} {b'}\partial_{\theta}\left[ \frac {A_{2}}{c^2}\left(q_{\theta \theta}^{\prime}-\frac {\Gamma} 2 \eta_{\theta \theta}^{\prime}\right)+\frac{1}{\rho_1} q^{\prime} \eta^{\prime}
+\left(\frac{\gamma_1}{2}-\frac{\Gamma}{2\rho_1}\right) \eta^{\prime 2}\right]
+ \mathcal { O } \left( \varepsilon ^ { 4 } \right), \nonumber
\end{align}
and by \eqref{3.17}, the above equation can be written as
\begin{align}\label{3.21}
q' _ { \theta }& = \frac{\rho_1} {b'} ( c - \kappa+ \frac{\Gamma b'}{2\rho_1}) \eta' _ { \theta } - \varepsilon^{2}\frac{\rho_1c} {b'}\partial_{X}\left((\kappa-\frac{\Gamma b'}{2\rho_1}+c_{1})\eta^{\prime}\right)\nonumber\\
&-\varepsilon^{2}\frac{\rho_1} {b'}\partial_{\theta}\left[ \frac {A_{2}}{c^2}(\frac{\rho_1}{b'}c_{1}-\frac {\Gamma} 2) \eta_{\theta \theta}^{\prime}
+\left(\frac{c_{1}}{b'}+\frac{\gamma_1}{2}-\frac{\Gamma}{2\rho_1}\right) \eta^{\prime 2}\right]
+ \mathcal { O } \left( \varepsilon ^ { 4 } \right).
\end{align}
Tedious calculations and some algebraic manipulations yield that the equation \eqref{3.20} can be written with $q$ excluded as
\begin{align}\label{3.22}
&c\eta'_{X}+\left( c_{X}-\frac {c_1c} {2b'}(\frac{b'}{c_{1}})_X \right) \eta'
 + \frac {A_{2}}{c^3}\frac{\rho_{1} c_{0}^{2}}{2c_{1}b'}\eta'_{\theta\theta\theta}\nonumber\\
&\quad +\frac 1 {2c c_{1}}\left(\frac{b'}{\rho_{1}}\left(\rho_1 \gamma_1^{2}-\rho_2\gamma_{2}^{2}\right)+3c_{0}
\left(\frac{c_{0}}{b'}+\gamma_{1}\right)\right)
\eta'\eta'_{\theta}=0,
\end{align}
where the relation \eqref{3.18} is used. This equation is a KdV-type equation \cite{KdV}, whose coefficients depend on functions of the slowly varying variable $X$.
\begin{remark}
Setting $\tilde{\theta}=c \theta$, then the equation \eqref{3.22} is reformed as
\begin{align}\label{3.23}
&c\eta'_{X}+\left( c_{X}-\frac {c_1c} {2b'}(\frac{b'}{c_{1}})_X \right) \eta'
 + A_{2}\frac{\rho_{1} c_{0}^{2}}{2c_{1}b'}\eta'_{\tilde{\theta}\tilde{\theta}\tilde{\theta}}\nonumber\\
&\quad +\frac 1 {2 c_{1}}\left(\frac{b'}{\rho_{1}}\left(\rho_1 \gamma_1^{2}-\rho_2\gamma_{2}^{2}\right)+3c_{0}
\left(\frac{c_{0}}{b'}+\gamma_{1}\right)\right)
\eta'\eta'_{\tilde{\theta}}=0.
\end{align}
In the special case $b'=h_1$ (i.e., taking $\beta=0$), the coefficients of the equation \eqref{3.23} recover the ones of the KdV equation in \cite{CoI19} for fluids with flat bottom, where a moving frame of reference and a scaling transformation are needed.
\end{remark}
\vspace{0.5cm}
\noindent {\bf Acknowledgements.}
The work of Fan is partially supported by a NSF of Henan Province of China Grant No. 222300420478 and the NSF of Henan Normal University Grant No. 2021PL04. The work of Gao is partially supported by NSFC No. 12171084 and the fundamental Research Funds for the Central Universities No. 2242022R10013..

\end{spacing}

\begin{thebibliography}{99}

\bibitem{BB1}{\small \textsc{T.B. Benjamin, T.J. Bridges}, Reappraisal of the Kelvin-Helmholtz problem. Part 1. Hamiltonian structure, {\it J. Fluid Mech.}, {\bf 333} (1997) 301-325.}

\bibitem{BB2}{\small \textsc{T.B. Benjamin, T.J. Bridges}, Reappraisal of the Kelvin-Helmholtz problem. part 2. interaction of the Kelvin-Helmholtz, superharmonic and Benjamin-Feir instabilities, {\it J. Fluid Mech.}, {\bf 333} (1997) 327-373.}

\bibitem{CC}{\small \textsc{W. Choi and R. Camassa}, Weakly nonlinear internal waves in a two-fluid system, {\it J. Fluid Mech.}, {\bf 313} (1996) 83-103.}

\bibitem{CC1}{\small \textsc{W. Choi and R. Camassa}, Fully nonlinear internal waves in a two-fluid system, {\it J. Fluid Mech.}, {\bf 386} (1999) 1-36.}

\bibitem{Co}{\small \textsc{A. Constantin}, {\it Nonliear Water Waves with Applications to Wave-Current Interactions and Tsunamis,} volume 81 of CBMS-NSF Conference Series in Applied Mathematics, SIAM, Philadelphia, (2011).}

\bibitem{CoI15}{\small \textsc{A. Constantin and R. I. Ivanov}, A Hamiltonian approach to wave-current interactions in two-layer fluids, {\it Phys. Fluids}, {\bf 27} (2015) 08660.}

\bibitem{CoI19}{\small \textsc{A. Constantin and R. I. Ivanov}, Equatorial wave-current interactions, {\it Commun. Math. Phys.}, {\bf 370} (2019) 1-48.}

\bibitem{CoIM}{\small \textsc{A. Constantin, R. I. Ivanov and C. I. Martin},  Hamiltonian formulation for wave-current interactions in stratified rotational flows, {\it Arch. Ration. Mech. Anal.}, {\bf 221} (2016) 1417-1447.}

\bibitem{CoIP}{\small \textsc{A. Constantin, R. I. Ivanov and E. M. Prodanov},  Nearly-Hamiltonian structure for water waves with constant vorticity, {\it J. Math. Fluid Mech.}, {\bf 9} (2007) 1-14.}

\bibitem{CoJ15}{\small \textsc{A. Constantin and R. S. Johnson}, The dynamics of waves interacting with the Equatorial Undercurrent, {\it  Geophys. Astrophys. Fluid Dyn. }, {\bf 109} (2015) 311-358.}

\bibitem{CoJ2017}{\small \textsc{A. Constantin and R. S. Johnson}, A nonlinear, three-dimensional model for ocean flows, motivated by some observations of the pacific equatorial undercurrent and thermocline, {\it Physics of Fluids}, {\bf 29} (2017) 056604 .}

\bibitem{CoM}{\small \textsc{A. Constantin and S. G. Monismith}, Gerstner waves in the presence of mean currents and rotation, {\it J. Fluid Mech.}, {\bf 820} (2017) 511-528.}

\bibitem{C1}{\small \textsc{A. Compelli}, Hamiltonian formulation of 2 bounded immiscible media with constant non-zero vorticities and a common interface, {\it Wave Motion}, {\bf 54} (2015) 115-124.}

\bibitem{C2}{\small \textsc{A. Compelli}, Hamiltonian approach to the modeling of internal geophysical waves with vorticity, {\it Monatsh. Math.}, {\bf 179} (2016) 509-521.}

\bibitem{CI}{\small \textsc{A. C. Compelli and R. I. Ivanov}, Hamiltonian approach to internal wave-current interactions in a two-media fluid with a rigid lid, {\it Pliska Stud. Math. Bulgar.}, {\bf 25} (2015) 7-18.}

\bibitem{CI0}{\small \textsc{A. C. Compelli and R. I. Ivanov}, On the dynamics of internal waves interacting with the Equatorial Undercurrent, {\it J. Nonlinear Math. Phys.}, {\bf 22} (2015) 531-539.}

\bibitem{CI1}{\small \textsc{A. C. Compelli and R. I. Ivanov}, The dynamics of flat surface internal geophysical waves with currents, {\it J. Math. Fluid Mech.}, {\bf 19} (2017) 329-334.}

\bibitem{CIMT}{\small \textsc{A. C. Compelli, R. I. Ivanov, C. I. Martin and M. D. Todorov}, Surface waves over currents and uneven bottom, {\it Deep-Sea Research Part II}, {\bf 160} (2019) 25-31.}

\bibitem{CIT}{\small \textsc{A. C. Compelli, R. I. Ivanov and M. D. Todorov}, Hamiltonian models for the propagation of irrotational surface gravity waves over a variable bottom, {\it Phil. Trans. R. Soc. A}, {\bf 376} (2017): 20170091.}

\bibitem{CG}{\small \textsc{W. Craig and M. Groves}, Normal forms for waves in fluid interfaces, {\it Wave Motion}, {\bf 31} (2000) 21-41.}

\bibitem{CGK}{\small \textsc{W. Craig, P. Guyenne and H. Kalisch}, Hamiltonian long wave expansions for free surfaces and interfaces, {\it Comm. Pure Appl. Math.}, {\bf 58} (2005) 1587-1641.}

\bibitem{CGNS}{\small \textsc{W. Craig, P. Guyenne, D. P. Nicholls and C. Sulem}, Hamiltonian long-wave expansions for water waves over a rough bottom, {\it Proc. R. Soc. A}, {\bf 461} (2005) 839-873.}

\bibitem{CGS}{\small \textsc{W. Craig, P. Guyenne and C. Sulem}, Coupling between internal and surface waves, {\it Nat. Hazards}, {\bf 57} (2011) 617-642.}

\bibitem{CuI}{\small \textsc{J. Cullen and R. I. Ivanov}, On the intermediate long wave propagation for internal waves in the presence of currents, {\it European Journal of Mechanics / B Fluids}, {\bf 84} (2020) 325-333.}

\bibitem{D}{\small \textsc{S. Dai}, Interactions between two pairs of the solitary waves in a two-layer fluid, {\it Science in China Ser. A}, {\bf 26} (1983) 1007-1017 (in Chinese).}

\bibitem{DCDPG}{\small \textsc{A. De Bouard, W. Craig, O. D\'{\i}az-Espinosa, P. Guyenne and C. Sulem}, Long wave expansions for water waves over random topography, {\it Nonlinearity}, {\bf 21} (2008) 2143-2178.}

\bibitem{IM}{\small \textsc{D. Ionescu-Kruse and C. I. Martin}, Periodic equatorial water flows from a Hamiltonian perspective, {\it J. Diff. Eqs.}, {\bf 262} (2017) 4451-4474.}

\bibitem{I}{\small \textsc{R. I. Ivanov}, Hamiltonian model for coupled surface and internal waves in the presence of currents, {\it Nonlinear Anal.: Real World Appl.}, {\bf 34} (2017) 316-334.}

\bibitem{IMT}{\small \textsc{R. I. Ivanov, C. I. Martin and M. D. Todorov}, Hamiltonian approach to modelling interfacial internal waves over variable bottom, {\it Physica D: Nonlinear Phenomena}, {\bf 433} (2022) 133190.}

\bibitem{Jo}{\small \textsc{R. S. Johnson}, On the development of a solitary wave moving over an uneven bottom, {\it Proc. Camb. Phil. Soc.}, {\bf 73} (1973)  183-203.}

\bibitem{Jo1}{\small \textsc{R. S. Johnson}, On an asymptotic solution of the Korteweg-de Vries equation with slowly varying coefficients, {\it J. Fluid Mech.}, {\bf 60} (1973)  813-824.}

\bibitem{KdV}{\small \textsc{D. J. Korteweg and G. Vries}, On the change of form of long waves advancing in a rectangular channel and on a new type of long stationary wave, {\it Philos. Mag.}, {\bf 39} (1895) 422-443.}

\bibitem{LDZ}{\small \textsc{D. Lu, S. Dai and B. Zhang},  Hamiltonian formulation of nonlinear water waves in a two-fluid system, {\it Appl. Math. Mech. -Engl. Ed.}, {\bf 20} (1999) 343-349.}

\bibitem{M}{\small \textsc{C. I. Martin}, A Hamiltonian approach for nonlinear rotational capillary-gravity water waves in stratified flows, {\it Discrete Continuous Dynamical Systems}, {\bf 37} (2017) 387-404.}

\bibitem{Wa}{\small \textsc{E. Wahl\'{e}n}, Hamiltonian long wave approximations of water waves with constant vorticity, {\it Phys. Lett. A.}, {\bf 372} (2008) 2597-2602.}

\bibitem{Za}{\small \textsc{V. E. Zakharov}, Stability of periodic waves of finite amplitude on the surface of a deep fluid, {\it J. Appl. Mech. Tech. Phys.}, {\bf 4} (1968) 190-194.}

\bibitem{ZP}{\small \textsc{H. Zhou and D. Piao}, Hamiltonian long wave expansions for internal waves over a periodically varying bottom, {\it Appl. Math. Mech. -Engl. Ed.}, {\bf 29} (2008) 745-756.}


\end{thebibliography}
\end{document}